\documentclass[preprint2]{emulateapj}
\usepackage{amsmath}
\usepackage{enumerate}
\usepackage{hyperref}
\usepackage{color}
\usepackage{longtable}

\shorttitle{GMAs in M100 Seen by ALMA}
\shortauthors{Pan et al.}

\begin{document}


\title{Variation in GMC Association Properties Across Bars, Spiral Arms, Inter-arms, and Circumnuclear Region of M100 (NGC 4321) extracted from ALMA Observations}



\author{Hsi-An Pan}
\affil{Academia Sinica, Institute of Astronomy \& Astrophysics (ASIAA), P.O. Box 23-141, Taipei 10617, Taiwan}\email{hapan@asiaa.sinica.edu.tw}
\author{Nario Kuno}
\affil{Faculty of Pure and Applied Sciences, University of Tsukuba, 1-1-1 Tennoudai, Tsukuba, Ibaraki 350-8577, Japan}
\affil{Center for Integrated Research in Fundamental Science and Engineering, University of Tsukuba, Tsukuba, Ibaraki 305-8571, Japan}








\begin{abstract} 
We study the physical properties of giant molecular cloud associations (GMAs) in M100 (NGC 4321) using the ALMA Science Verification feathered (12-m+ACA) data in $^{12}$CO (1-0). To examine the environmental dependence of GMA properties, GMAs are classified based on their locations in the various environments as circumnuclear ring (CNR), bar, spiral, and inter-arm GMAs. The CNR GMAs are massive and compact, while the inter-arm GMAs are diffuse with low surface density. GMA mass and size are strongly correlated, as suggested by Larson (1981). However, the diverse power-law index of the relation implies that the GMA properties are not uniform among the environments. The CNR and bar GMAs show higher velocity dispersion than those in other environments. We find little evidence for a correlation between GMA velocity dispersion and size, which indicates that the GMAs are in diverse dynamical states. Indeed, the virial parameter of GMAs spans nearly two orders of magnitude. Only the spiral GMAs are in general self-gravitating. Star formation activity of the GMAs decreases in order over the CNR, spiral, bar, and the inter-arm GMAs. The diverse GMA and star formation properties in different environments lead to variations in the Kennicutt-Schmidt relation. A combination of multiple mechanisms or gas phase change is necessary to explain the observed slopes. Comparisons of GMA properties acquired with the use of the 12-m-array observations with those from the feathered data are also presented. The results show that the missing flux and  extended emission cannot be neglected for the study of environmental dependence.
\end{abstract}





\section{Introduction}

The physical properties of giant molecular clouds (GMCs), which are the seeds of star formation, determine whether or not a star can form. Therefore, an understanding of the factors that determine the properties of GMCs is crucial to understand the star formation process for,  e.g., why stars are formed in certain regions, and what controls the efficiency of star formation of a GMC.

Mounting evidence suggests that the cloud properties depend on both local and global environments. Galactic cloud properties were studied in the 1980s by \cite{Lar81}  and \cite{Sol87}, who suggested that clouds are gravitationally bound structures supported by turbulence against self-gravity. These properties lead to the implication that the surface density of clouds is approximately constant. Accordingly, \cite{Lar81}  and \cite{Sol87} proposed that the cloud properties of the Milky Way are decoupled from their environments, namely, cloud properties are universal. However, further studies have revealed that cloud properties can be affected by local environments and conditions such as the ambient pressure \citep{Bli04,Hug13,Mei16}  and star formation and feedback (e.g., supernovae, radiation pressure, and ionization) in the cloud \citep{Mck89,Wol95,Tas11}. In addition, \cite{Hey01}  have suggested that the cloud properties depend on larger, galactic-scale environments. They found that many clouds in the outer Galactic disc have low mass and are not gravitationally bound. In other words, the outer-disc clouds do not share the same properties as the inner-disc clouds observed by \cite{Lar81} and \cite{Sol87}.

In recent years, extragalactic observations of clouds and cloud associations  have covered large areas in nearby galaxies. Extragalactic observations provide a perfect perspective for galactic-scale environments that are difficult to obtain from the edge-on Galactic observations. These observations of nearby galaxies redefine our understanding of galactic-scale environments such as galactic structures (bars, spiral arms, inter-arm regions) \citep{Hug13,Col14,Pan15a}, regions of shear arising from galaxy rotation \citep{Kod09,Mei13,Miy14}, and galaxy types (dwarf, early-, and late-type galaxies) \citep{Ler08,Hug13,Thi14} on the regulation of cloud properties. The importance of galactic-scale environments is also supported by galaxy simulations \citep{Fuj14,Ren15}.

However, most of the observational studies have concentrated on cloud properties alone with high physical resolution   ($<$ 100 pc), but  there is a lack of   comparison between cloud and star formation  to fill in the big picture. 
Time-averaged quantities such as the star formation rate (SFR) require measurements over larger scales to sample the full stellar evolution of individual regions \citep{Cal12,Kru14}. Moreover,  stars can decouple from their parent clouds either because the clouds dissociate or disperse by feedback, or because galactic dynamics makes the orbits of the stars and gas diverge \citep{Ono10,Kru14}.   
Thus, information linking molecular gas and star formation to galactic-scale environments is mainly available at sub-kpc resolutions. 
  In addition,  observational studies of galactic-scale environments have mostly focused on the comparisons of cloud properties between spiral and inter-arm regions or between central and disc regions due to the resolution limit and inclination of the galactic disc.

In this study, using the released Science Verification (SV) data of ALMA, we consider the galactic-scale environmental effects on molecular gas and star formation activity in M100 (NGC 4321) with object-based analysis and a finer classification of galactic environments for the first time. The low inclination of M100 affords a perfect perspective of galactic structures, allowing us to compare the GMA properties between various environments (circumnuclear region, bars, spiral arms, and inter-arm regions) at once. M100 is located at the outskirts of the Virgo Cluster. We use the distance obtained using Cepheids, 14.3 Mpc, yielding a linear scale of 1$\arcsec$ $=$ 69 pc \citep{Fre01}. The adopted inclination and the position angle of the major axis are 27$^{\circ}$ and 153$^{\circ}$, respectively \citep{Kna93}.

The morphology of M100 has been classified as SABbc by \cite{Vau91}. The galaxy is highly structured. Two spiral arms emerge from the primary bar, which has a length of $\sim$ 4.5 kpc from the galactic center. The primary bar wraps around the galactic nucleus, forming a circumnuclear ring (CNR) with a radius of $\sim$ 1 kpc. A nuclear (secondary) bar aligned parallel to the primary bar is found in the CNR \citep{Sak95,Gar98}. The nucleus of M100 has been classified as HII/LINER by \cite{Ho97}. The galaxy has two close companions within $\sim$ 6$\arcmin$: NGC 4322 to the north, and NGC 4328 to the east. HI observations show a tail extending to the southwest, thus suggesting recent or ongoing interaction \citep{Kna93}. The interaction scenario is also supported by the asymmetric polarized emission of M100 \citep{Wez12,Vol13}. Although M100 has interacting companions, the spiral arms of M100 appear to be symmetric \citep{Elm11}, and thus, it can be considered to represent the normal features of isolated galaxies.

The paper is organized as follows. In Section \ref{sec_data}, we present a summary of the ALMA SV feathered data. Section \ref{Sec_Obj} discusses the identification of GMAs and the calculation of their SFRs.  In Section \ref{Subsec_gal_env}, we compare the GMA properties formed in different galactic environments. Discussion and comparisons with previous studies are presented in   Section \ref{sec_discussion}. In Section \ref{sec_obs_effect}  we compare GMAs identified in the feathered data with those identified in the 12-m-alone observation to examine the effect of missing flux on the interpretation of the environmental dependence of GMAs. The key results are summarized in Section \ref{Sec_summary}.

\section{Data}
\label{sec_data}
We used the processed 12-m + Atacama Compact Array (ACA, ACA $=$ 7-m + total power) archival ALMA SV data at Band 3 (115 GHz; $^{12}$CO (1-0)) on M100. The full description of the data reduction and the combination of the three observing modes can be found in the Common Astronomy Software Applications (CASA 4.3) guide for this specific dataset\footnote{\url{https://casaguides.nrao.edu/index.php/M100_Band3}}. We provide a brief summary of the data here. The ALMA 12-m, 7-m array, and total power observations were carried out in 2011, 2013, and 2014, respectively. The beamsize of the 12-m-alone observation was 3.46$\arcsec$ $\times$ 2.37$\arcsec$, that of the  7-m-alone observation was 12.72$\arcsec$ $\times$ 10.12$\arcsec$, and that of the total power observation was 56.9$\arcsec$ $\times$ 56.9$\arcsec$. In our study, the 7-m and 12-m interferometric data are combined first. The combined interferometric map is processed using the CLEAN algorithm with a robust $=$ 0.5 weighting (Briggs) of the visibilities. The resulting image is feathered with the total power image to recover the  extended emission. The total flux of the combined interferometric map is $\sim$ 1400 Jy, and it increases to $\sim$ 3000 Jy (after primary-beam correction) upon adding the total power data to the interferometric data. The final image covers an area of $\sim$ 200$\arcsec$ $\times$ 200$\arcsec$ (14 $\times$ 14 kpc). The spatial resolution of the map is 3.87$\arcsec$ $\times$ 2.53$\arcsec$ ($\sim$ 267 $\times$ 174 pc) and P.A. of -89.51$^{\circ}$. The CO data are binned into 5.023 km s$^{-1}$. The rms noise ($\sigma_{\mathrm{rms}}$) of the data cube is $\sim$ 0.015 Jy beam$^{-1}$, corresponding to a molecular mass sensitivity of $\sim$10$^{5}$ M$_{\sun}$ per beam per 5 km s$^{-1}$ channel.

\begin{figure}
\epsscale{1.1}
\plotone{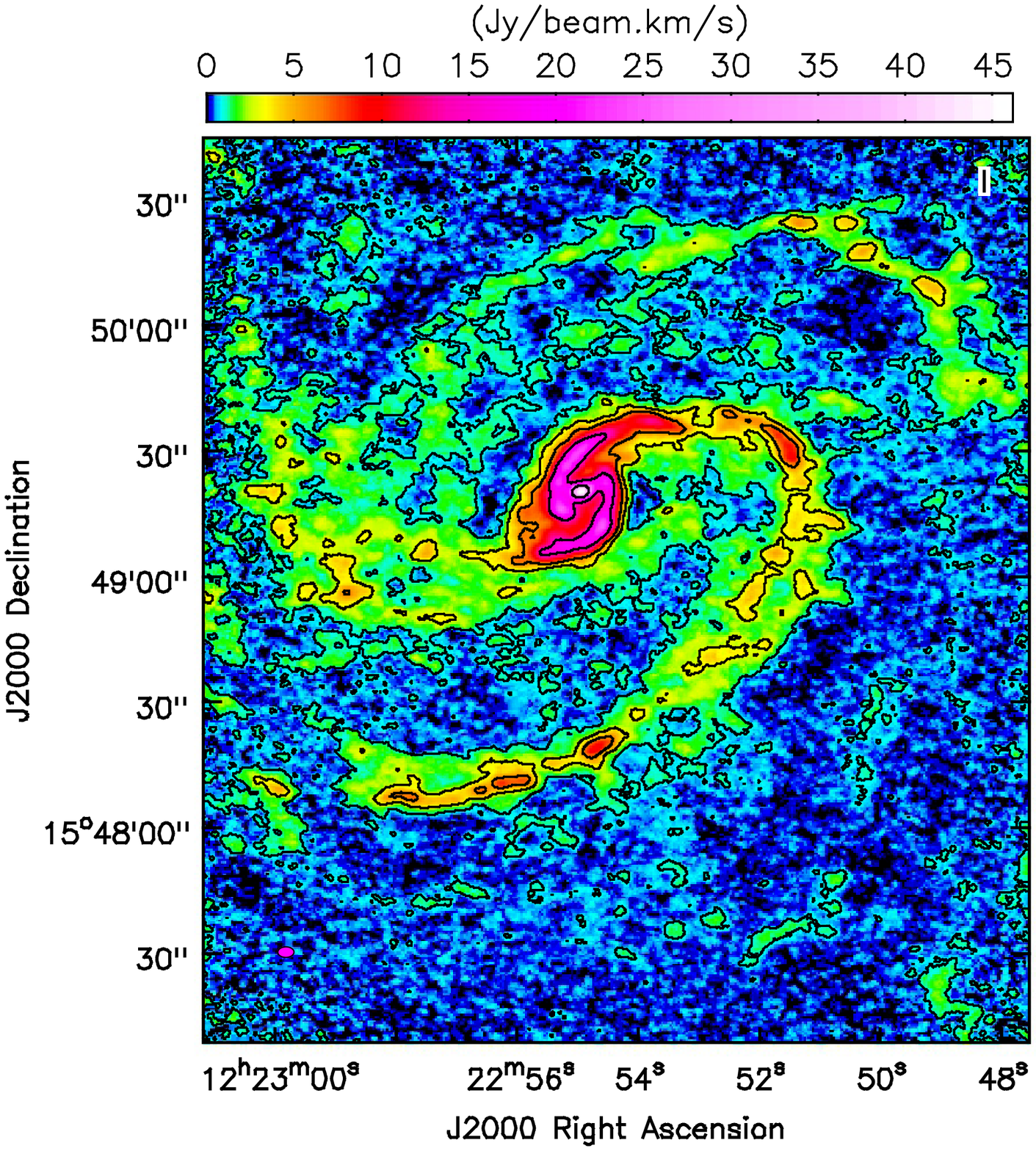}
\caption{Integrated intensity map of ALMA $^{12}$CO (1--0) feathered (12m$+$ACA) data. The contour levels are in step of 3, 10, 20, 50, and 100 $\sigma$, where 1 $\sigma$ $=$ 0.3 Jy beam$^{-1}$ km s$^{-1}$. The beam size  is plotted at the lower-left corner in magenta color.
\label{FIG_SDSSCO}}
\end{figure}

\begin{figure}
\epsscale{1}
\plotone{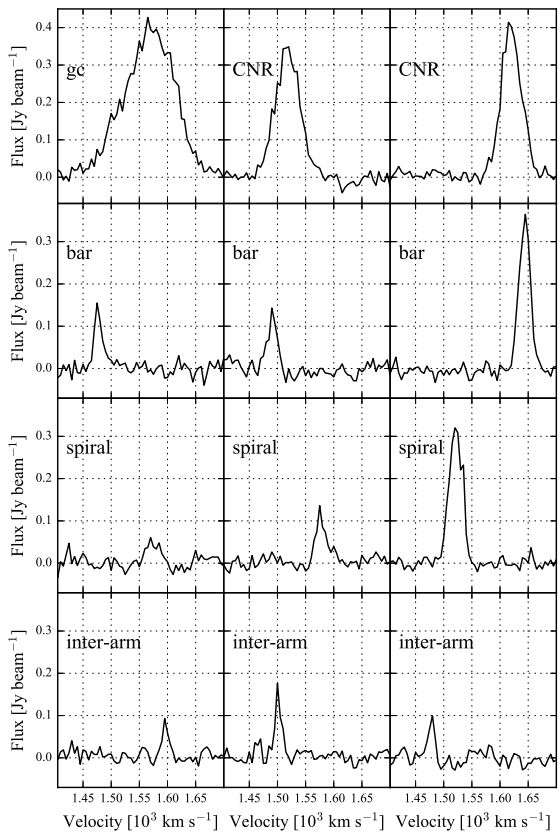}
\caption{Representative spectral lines  at the galactic center (gc), circumnuclear ring (CNR), bar, spiral, and inter-arm regions. The noise level in the spectral data cube is $\sim$ 0.015 Jy beam$^{-1}$.
\label{FIG_Spec}}
\end{figure}

Figure \ref{FIG_SDSSCO} presents the integrated intensity map of $^{12}$CO (1-0).
The morphology of molecular gas is characterized by several components.
There is a strong concentration of molecular gas at the galactic center. Its shape is similar to the beam, presumably representing  an unresolved source. The  unresolved source was detected at a signal-to-noise ratio (S/N) of $\sim$ 30 as shown in the spectral line in Figure \ref{FIG_Spec}. The central concentration is surrounded  by a ring-like structure  formed by two spiral arms  wrapping up at a radius of $\sim$ 15$\arcsec$. The ring-like structure was detected at S/N $\approx$ 20 -- 30. The middle and right panels of Figure \ref{FIG_Spec} display two representative spectral lines at the north and south central spiral arms, respectively.
The central spiral arms do not extend all the way to the central source, but are connected with a nuclear bar that extends  to the  center.
The primary bar emerges from another end of the central spiral arms. The western bar is   narrow with relatively strong CO emission, while the CO emission in the eastern bar is more diffuse and weaker.  
 A similar gas distribution is  seen in the spiral arms connecting  the two sides of the bar.
CO emission along the southern spiral arm is uneven,   with a clumpy hierarchical structure with strong emission. The bar and spiral regions also exhibit high S/N in the range from 3 -- 20, as can be observed from the middle two rows of Figure \ref{FIG_Spec}.
It is remarkable that many inter-arm regions were  detected with S/N $>$ 3.
Most of the inter-arm emission in the integrated map can pass the criteria to be a cloud structure, allowing us to gain insight into gas properties in these largely unexplored regions.

\section{Identification of GMAs and Their  HII Regions}
\label{Sec_Obj}

Many studies have used an approach that identifies individual molecular clouds and then measures their properties in the spectral data cube (position-position-velocity space).  These identification algorithms often make several assumptions for segmentation of adjacent clouds and property measurements. \cite{Ler16} have shown that the cloud properties measured by such object-based methods generally show good agreement with the pixel-based  approach  in the integrated intensity map (position-position space) for low-inclination galaxies. Nonetheless, the object-based method is still useful in reducing the line-of-sight projection effect, eliminating diffuse and low S/N emission, and showing the cloud axis-ratio and orientation that are believed to have implications on the influence from galactic-scale structures. Therefore, we adopt the object-based analysis in this work utilizing a widely used identification algorithm.

\subsection{GMA identification}
\label{sec_gmas_cprops}

The molecular cloud structures are identified by the cloud property algorithm CPROPS \citep{Ros06}. The CPROPS process begins with the masking of the emission with a high S/N ($m$ $\times$ $\sigma_{\mathrm{rms}}$), thereby picking out the pixels with intensity significantly higher than the background. In CPROPS, parameter $m$ $\times$ $\sigma_{\mathrm{rms}}$ is estimated from the median absolute deviation (MAD) of each spectrum. CPROPS subsequently extends this mask to the user-defined lowest signal-to-noise ratio ($e$  $\times$  $\sigma_{\mathrm{rms}}$), which outlines the boundary of significant emission ($T_{\mathrm{edge}}$). 
To ensure consistency with previous studies, we set $m$ $=$ 4 and $e$ $=$  1.5. After the  regions of significant emission are identified, CPROPS searches these to locate separate peaks. The search for the peaks is performed within a cube with a box with dimensions of $\sim$ 660 pc $\times$ 660 pc $\times$ 15 km s$^{-1}$, corresponding to three times the beam and channel width, respectively. 
 If only one peak is found within an emission region, then CPROPs labels the region as  a discrete object with \emph{observed boundary} $T_{\mathrm{edge}}$ and measure  its properties. 
On the other hand, if multiple peaks are found within an emission region, CPROPS proceeds to verify each maximum's independence using a modified watershed algorithm, where the maximum is required to lie at least 2$\sigma_{\mathrm{rms}}$ above the \emph{merge level} with  another  maximum. In other words, only the emission that is \emph{uniquely} associated with a maximum is given an assignment (i.e. only  emission that is above  merge level). The remainder of the emission is considered to be in the watershed.  In this case, the \emph{observed boundary}  has a brightness greater than  $T_{\mathrm{edge}}$.
Because of the limitation of the map resolution, we disable this procedure so that the local maxima are forcibly separated and form individual objects. We determined that it is better to mark the objects that visual inspection would suggest are independent structures and to prevent the local maxima merging into larger objects that outline the large-scale galactic structures (e.g., a portion of a curved spiral arm-like structure). Around 60\% of the flux in the data cube is assigned to objects.

Since the GMCs are not resolved with our resolution, hereafter, we refer to the structures identified by CPROPS as giant molecular cloud associations (GMAs). Note that our angular resolution of $\sim$ 3$\arcsec$ corresponds to $\sim$ 215 pc, which is high enough to isolate (but not resolve) single (or at most a few)  GMCs, given that the typical separation of  GMCs in  Milky-Way-like barred spiral galaxies  is a few 100 pc to kpc \citep{Sol87,Kod06}. Therefore, it is reasonable to assume that the measured GMA properties can represent the local GMC(s) properties and local star formation activity.

Only highly reliable GMAs are adopted in this work. GMAs found at the edge of the map or having peak intensity less than 5$\sigma_{\mathrm{rms}}$ are neglected. CPROPS performs deconvolution to correct the resolution on the GMA radius and velocity dispersion (cf. \S\ref{Subsec_gal_env}). If any deconvolution fails, the GMA is removed from our catalog. 
The final number of GMAs is 165, accounting for $\sim$ 55\% of the total flux in the data cube. In most cases, for a given position, only one GMA is found, and therefore, the projection effect along the line of sight is negligible. This is because the sub-kpc resolution captures the bulk molecular gas, which has a thickness close to that of the galactic disc ($\sim$100 - 150 pc). Only four GMAs are clearly separated along the velocity axis but overlap each other along the spatial axes. We also check the overlap between the edges of adjacent GMAs. The observed boundaries of the majority of GMAs are largely isolated; only a few GMAs exhibit a few pixels overlapped with adjacent GMAs. The 165 GMAs and their properties are listed  in the Appendix.

 CPROPS uses bootstrapping  of the assigned pixels  to estimate the uncertainties  of cloud properties. 
The bootstrapping method produces uncertainty in the measurement of the properties of a defined cloud, but neglects the uncertainty due to noise fluctuations or choice of algorithm (see Section 2.5 of \Citealt{Ros06} for the details). \cite{Col14} found that 50 bootstrapping measurements provide a reliable estimate of the uncertainty, and thus, this number is adopted in this work.

\begin{figure*}
\epsscale{1}
\plottwo{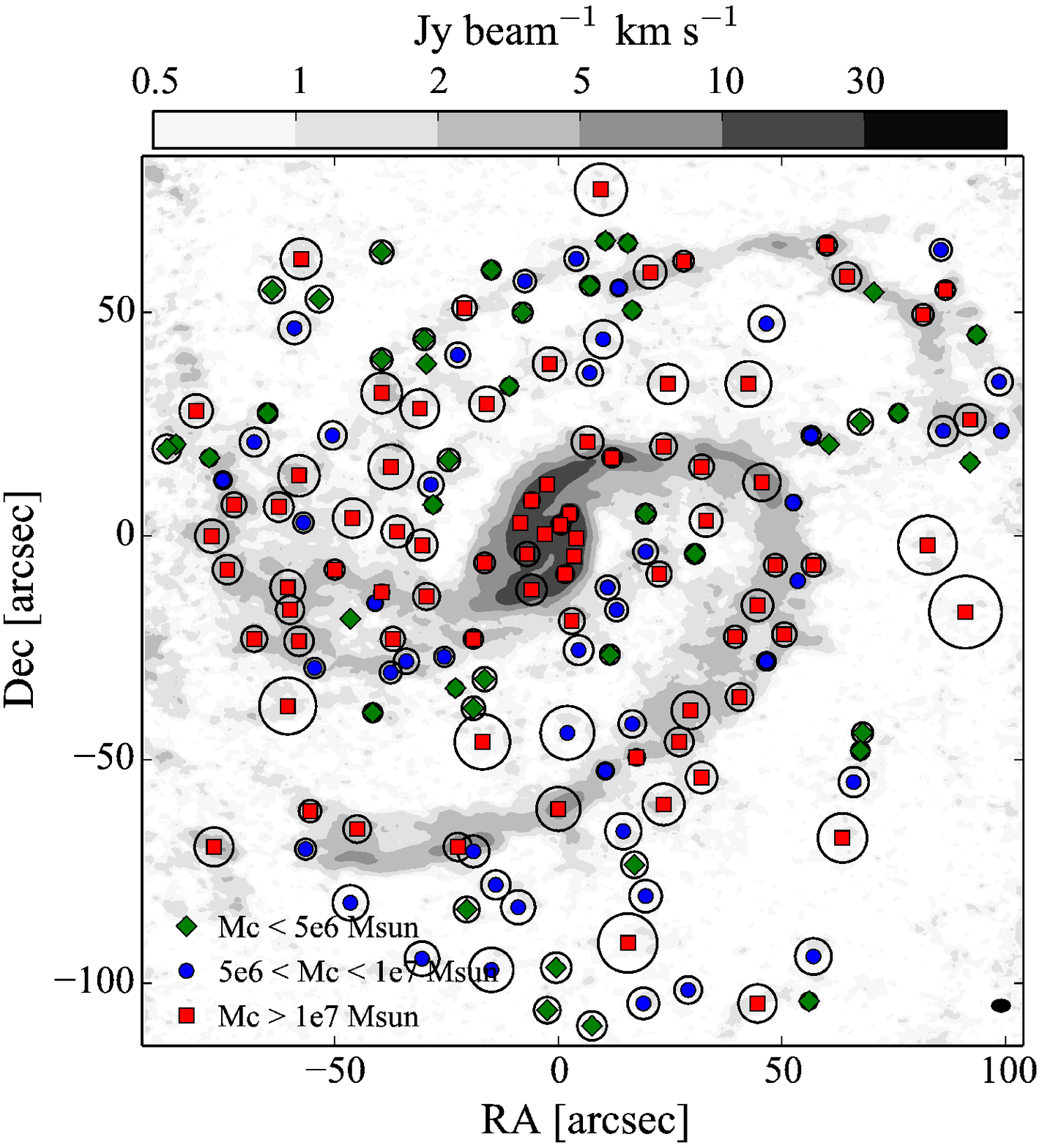}{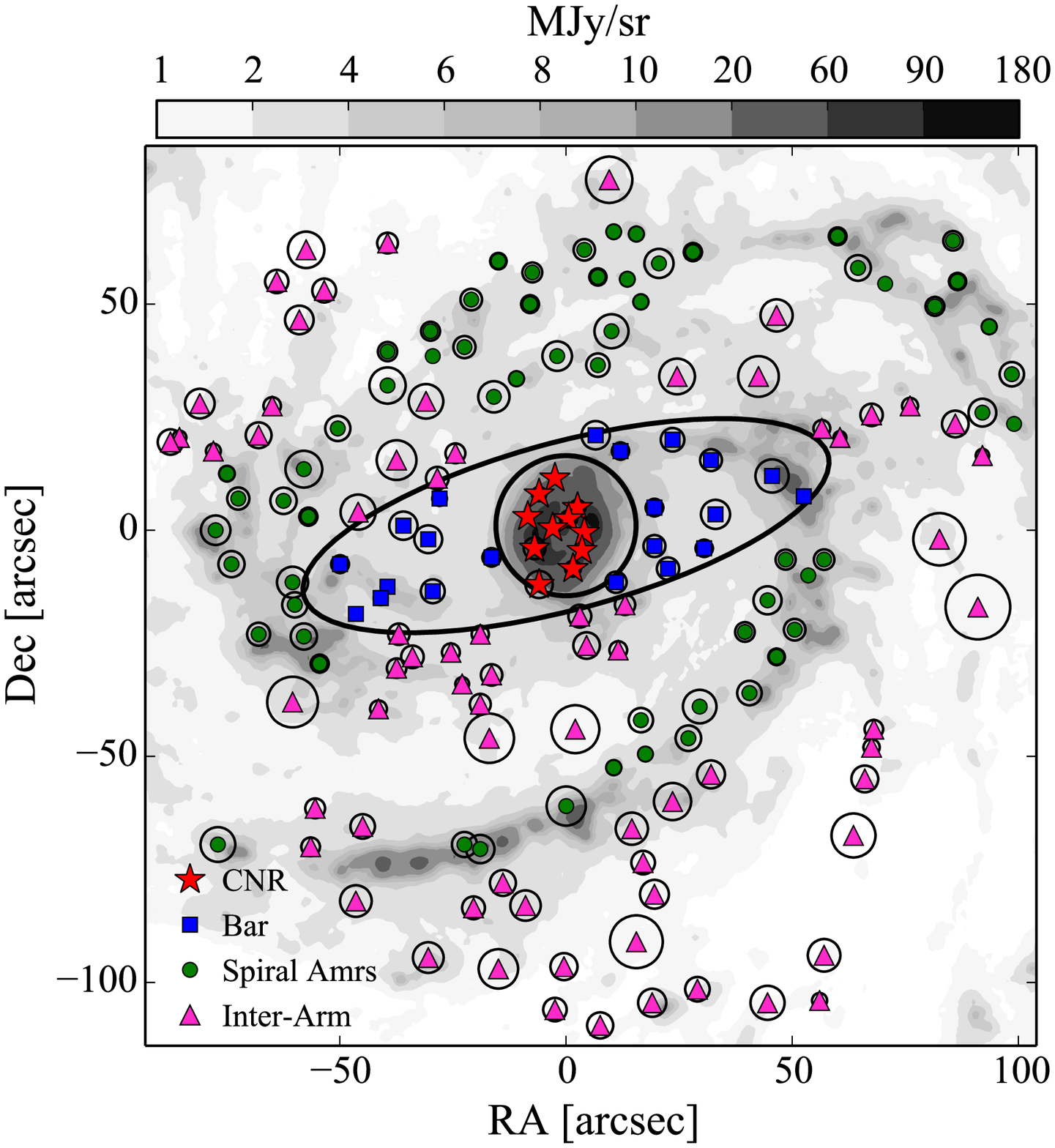}
\caption{Left: Spatial distribution of GMA mass ($M_{\mathrm{c}}$) and observed GMA boundary.  The $^{12}$CO integrated intensity map is presented with gray scale. We identify 165 reliable GMAs. Green diamonds, blue circles and red squares  denote GMA with $M_{\mathrm{c}}$  $<$ 5 $\times$ 10$^{6}$ M$_{\sun}$,  5 $\times$ 10$^{6}$ $<$ $M_{\mathrm{c}}$  $<$  10$^{7}$ M$_{\sun}$, and $M_{\mathrm{c}}$ $>$  10$^{7}$ M$_{\sun}$, respectively. The circles indicate the \emph{average} radius of the major and minor axes of the observed GMA boundary. Right: Classification of GMAs based on their locations in the galaxy. The background image is Spitzer 8$\mu$m emission. The central circle and ellipse denote the area  of the CNR and bar, respectively. The CNR GMAs, bar GMAs, spiral GMAs, and inter-arm GMAs are indicated with red stars, blue squares, green circles, and magenta triangles.   The numbers of GMAs  are: 11 CNR GMAs, 21 bar GMAs, 62 spiral GMAs, and 71 inter-arm GMAs.
\label{FIG_Mom0_GMAs}}
\end{figure*}

The left panel in Figure \ref{FIG_Mom0_GMAs} shows the 165 GMAs superposed on the CO integrated intensity map. The green diamonds, blue circles, and red squares in the figure denote GMAs with mass $<$ 5 $\times$ 10$^{6}$ M$_{\sun}$,  5 $\times$ 10$^{6}$ $<$ mass  $<$  10$^{7}$ M$_{\sun}$, and  $>$  10$^{7}$ M$_{\sun}$, respectively (the derivation of mass is presented in \S\ref{Subsec_gal_env}). The black open circles in the figure indicate the average radius of the area of the GMAs’ observed boundaries\footnote{Note that the real observed boundary of GMAs are not spherical.}, thereby affording a rough idea of the size of the high-density region (strong CO emission) assigned to each GMA.

\subsection{HII Regions and SFR of GMAs}
\label{sec_calc_sfr}
 The star formation rate (SFR) of a GMA is calculated from the total flux of the HII regions enclosed within the high-density region of the GMA (observed boundary). This is to ensure that only the high-confidence HII regions that are associated with high-density regions are selected. As a result, the derived SFR may be a lower limit.

\begin{figure*}
\epsscale{1}
\plottwo{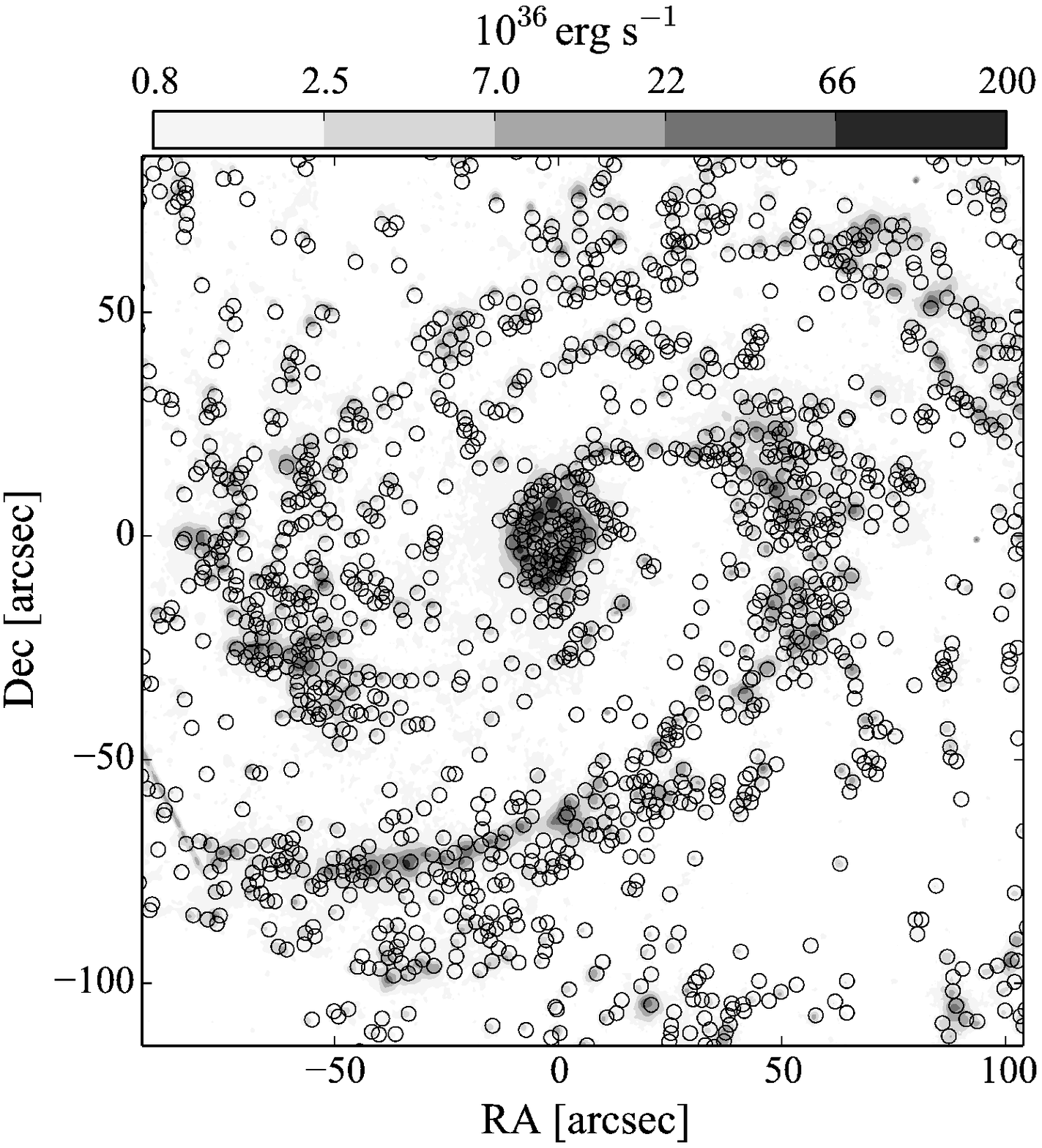}{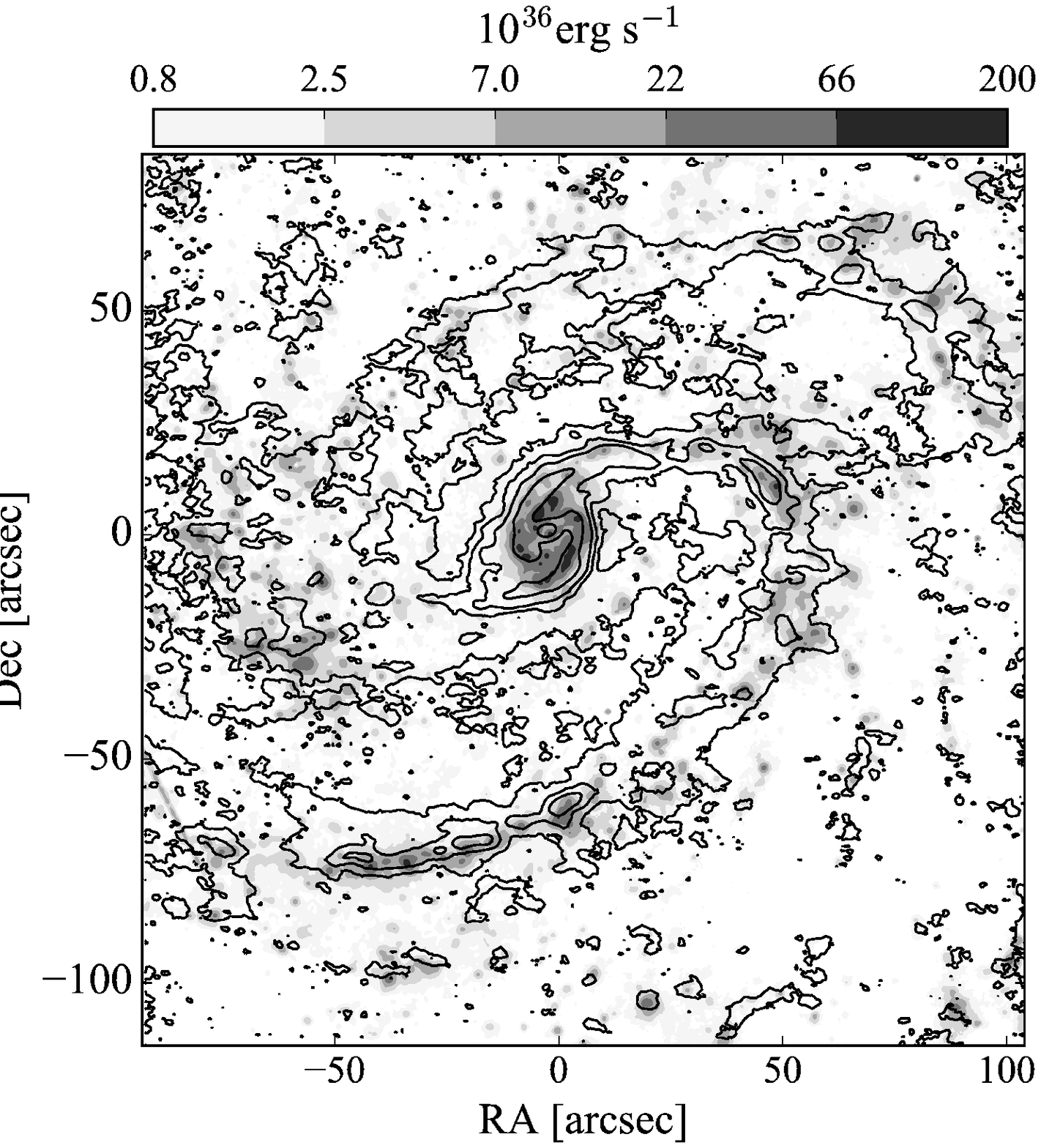}
\caption{Left: The continuum- and sky-background-subtracted H$\alpha$ image taken from \cite{Kna98}. The HII regions identified by  \cite{Kna98} are indicated with black circles. Note that the size of the circle do not represent the physical size of the HII regions. Right:  ALMA $^{12}$CO (1--0) image (contours)   overlaid on the H$\alpha$ image.  The contour levels are in step of 3, 10, 20, 50, and 100 $\sigma$, where 1 $\sigma$ $=$ 0.3 Jy beam$^{-1}$ km s$^{-1}$ (same as Figure \ref{FIG_SDSSCO}). 
\label{FIG_Halpha}}
\end{figure*}

The catalogue of HII regions is taken from \cite{Kna98}. H$\alpha$ observations were carried out with the 4.2-m William Herschel Telescope. The left panel in  Figure \ref{FIG_Halpha} shows the continuum- and sky-subtracted H$\alpha$ image in the color scale. From the image, \cite{Kna98}  catalogued 1948 HII regions, which are indicated by black circles in the figure. 
 The HII region should consist of at least nine  pixels with an intensity  at least 3 times the rms noise level of the local background. The flux of each HII region was determined by integrating counts within a circular aperture.  The estimated uncertainty in the flux of an HII region is about 10\%,  including the errors in the photometric calibration, the relative calibration of subimages, the sky-subtraction errors,  and  the uncertainty of galaxy distance.  The uncertainty in the flux of the HII regions is rather constant across the disc.

Stars can decouple from their parent clouds because either the clouds dissociate or disperse by feedback, or because galactic dynamics makes the orbits of the stars and gas diverge. The former case is unlikely to be resolved with our resolution \citep{Ono10,Kru14}, and thus, the formation and evolution of the HII regions should be associated with the surrounding GMA. In an attempt to examine the latter case, the right panel in Figure \ref{FIG_Halpha} compares the CO (contours) and H$\alpha$ (gray scale) emission. The H$\alpha$ emission is located at the leading side of the bar and spiral arms, but the peak emissions of CO and H$\alpha$ do not deviate from each other significantly. We also compared the observed cloud boundary (CO-bright, high-density regions), H$\alpha$ emission, and the defined HII regions visually (not shown in the paper), and found no significant offset between these populations with this resolution. Therefore, in this study, we assume that the GMAs and the HII regions coexist. Four GMAs that are distinctly separated along the velocity axis but overlapping with each other along the spatial axis are excluded in the calculation and discussion of star formation activity, but  retained for the analysis of GMA properties.

With this method, we find that about 17\% of the GMAs are not associated with any HII region. The areas of GMAs with and without HII regions lie in the range from 0.02 - 0.74 and 0.03 - 0.48 kpc$^{-2}$, respectively. The non-star-forming GMAs are not particularly small. Moreover, GMAs without HII regions are found everywhere in the galaxy, including the circumnuclear region, bar, spiral arms, and the inter-arm regions. Therefore, we propose that the absence of HII regions in these GMAs is not due to any obvious bias but likely a physical cause.

The SFR of a GMA is calculated by determining the sum of the SFRs of the HII regions within the GMA. For the SFR of each HII region, the calibration of \cite{Cal07} is used:
\begin{equation}
\mathrm{SFR(H\alpha )}=(5.3\times 10^{-42}L_{\mathrm{H}\alpha })\times 10^{A\mathrm{_{H\alpha}}/2.5}.
\end{equation}
Here, $L_{\mathrm{H}\alpha}$ denotes the observed luminosity of H$\alpha$ in erg s$^{-1}$, and $A\mathrm{_{H\alpha}}$ is the H$\alpha$ attenuation. 
The  $A\mathrm{_{H\alpha}}$ value derived by \cite{Pre07} is adopted for this work. \cite{Pre07} use the combination of mid-infrared (24$\mu$m) luminosity and H$\alpha$ luminosity of SFRs to derive $A\mathrm{_{H\alpha}}$. The 24$\mu$m luminosity is a tracer of obscured star formation owing to its long wavelength, and the uncorrected H$\alpha$ luminosity is a tracer of the unobscured portion. Parameter $A\mathrm{_{H\alpha}}$ can be estimated from the ratio of the total (obscured + unobscured) emission and the unobscured emission. The value of $A\mathrm{_{H\alpha}}$ $=$ 2.4 mag is adopted for the CNR region, while $A\mathrm{_{H\alpha}}$ $=$ 1.7 mag\footnote{A comparable value of $A\mathrm{_{H\alpha}}$ = 1.5 mag is derived from \cite{Won02} upon comparing the observed integrated H$\alpha$ flux of the galactic disc of M100 and that predicted by the thermal flux of the radio continuum.} is used for the bar and the spiral arms. \cite{Mar99}  determined that the extinction for the bar and disc HII regions are similar in a set of barred galaxies. Moreover, the $A\mathrm{_{H\alpha}}$ values of M100's HII regions fluctuate around 1.7 from $\sim$ 1.5 to 8 kpc \citep{Pre07}. For these reasons, we adopt the same $A\mathrm{_{H\alpha}}$ for the bar and the spiral regions.   The uncertainty in $A\mathrm{_{H\alpha}}$ is $\sim$ 0.8 mag, corresponding to a factor of 2 in SFR. The uncertainty arises from a variety of sources such as the geometry of the stars, dust, and gas, and the age and mass function of the embedded cluster. The inter-arm regions have  little, if any, dust  \citep{Kna96,Bec96}, and therefore, extinction correction is not applied to the inter-arm HII regions. For those GMAs with HII regions, SFR ranges between 2.1 $\times$ 10$^{-5}$ -- 0.3 M$_{\sun}$ yr$^{-1}$, with a median value 1.8 $\times$ 10$^{-3}$ M$_{\sun}$ yr$^{-1}$. 
The total SFR in the 165 GMAs is $\sim$ 2 M$_{\sun}$ yr$^{-1}$, accounting for $\sim$ 75\% of the global SFR derived from the H$\alpha$ + 24 $\mu$m data \citep{Wil09}.

\section{GMAs in Galactic Environments}
\label{Subsec_gal_env}
\subsection{GMA Classification Based on Galactic Environment}
\label{sec_classify}

GMAs are classified into four groups based on their locations in the galaxy. The classifications are visualized on the IRAC 8$\mu$m image in the right panel of Figure \ref{FIG_Mom0_GMAs}.

If a GMA is found to lie within the galactic radii of $r$ $<$ 15$\arcsec$ (1 kpc), we identify this GMA as a circumnuclear ring (CNR) GMA \citep{Sak95}. 
 Gas flow simulations of  M100 have shown that the ensemble of observations (optical, infrared, HI and CO maps) is best explained by two independent discs driven by two bars, where the two discs are dynamically decoupled. The inner disc, which has a form of CNR, has a radius of  15$\arcsec$ (1 kpc), while the outer disc is 8 -- 9 kpc. 
 Therefore,  GMAs located at $r$ $<$ 15$\arcsec$ are identified as CNR GMAs (central open circles  in the figure).   CNR GMAs are marked with red stars. The small open circles in the figure denote the observed average radius of each GMA (cf. \S\ref{sec_prop_MRSig}).  The radii of most of the CNR GMAs are smaller than the star signs.

Bar GMAs form in an elliptical region at the galactic center, as determined by \cite{She01}. The outermost bar isophote determined by \cite{She01} is marked with an open  ellipse in the right panel of Figure \ref{FIG_Mom0_GMAs}. This isophote was determined via fitting ellipses to optical and near-infrared images using the method of \cite{Reg97}. Bar GMAs are shown as blue squares in the figure. 

The spiral arms of M100 are isolated from the {\bf underlying stellar disc} with separate estimates from 3.6 $\mu$m, 4.5 $\mu$m and optical colour  by \cite{Ken11}. The azimuthal offset of the spiral arm ridge line between the wavelengths is $\sim$ 0-5$\arcsec$ along the arms.  The offset is caused by the uncertainty of the determination of the spiral arm ridge with different wavelengths.  The resolution of CO observations is comparable with the offset, and thus, we use the offset as the width of the spiral arms.
It is noteworthy that the northern and southern arms are not symmetric, e.g., the northern spiral arm is thicker along some segments due to the larger offset between the estimates and more diffuse infrared and optical emission distributions.  This may  be due to the influence of one or both companion galaxies \citep{Ken11}.
GMAs in the spiral pattern are classified as spiral GMAs. 
The rest of the GMAs in the disc are assigned as inter-arm GMAs. 
The spiral and inter-arm GMAs are marked with green circles and magenta triangles in the right panel of Figure \ref{FIG_Mom0_GMAs}, respectively. 
The numbers of GMAs classified according to each environment are: 11 CNR GMAs, 21 bar GMAs, 62 spiral GMAs, and 71 inter-arm GMAs.

\subsection{Comparison of GMA Properties between the Galactic Environments}
\label{sec_gmc_env_compare}

Figure \ref{FIG_Env_Properties} presents the normalized probability distribution of the GMAs and star formation properties for different environments. The color and symbols correspond to those in the right panel of Figure \ref{FIG_Mom0_GMAs}. The open symbols at the upper side of the panels indicate the median values. Error bars represent Poisson errors.

 To examine the existence of environmental dependence, the two-sample $t$-test  is applied to check whether the average difference of a GMA property between two environments is really significant. The derived $p$-value indicates the probability that the two environments are the same.  For example, $p$ = 0.01 implies that there is a 99\% chance of the two environments being  significantly different.
For each environment, we calculated the $p$-value with three other  environments separately.  These results are listed in Table \ref{TAB_p_value}.  The $p$-value  is interpreted in the following usual way:
$<$ 0.01 for \emph{highly significant} (two environments are significantly different),  0.01 -- 0.05 for   \emph{significant},  0.05 -- 0.1 for \emph{suggestive}, and $>$ 0.1 for \emph{non-significant} (two environments are likely the same).  For convenience, in Table \ref{TAB_p_value} the $p$-value are marked with $\times$, $\triangle$, $\lozenge$, and $\bigcirc$ for  highly significant difference, significant difference, suggestive of difference, and non-significant difference, respectively.

\subsubsection{Mass($M_{\mathrm{c}}$), Radius ($R_{\mathrm{c}}$), Surface Density ($\Sigma_{\mathrm{H_{2}}}$) and $M_{\mathrm{c}}$-$R_{\mathrm{c}}$ Relation}
\label{sec_prop_MRSig}
Since a GMA's outer regions are obscured by background noise, to determine the actual GMA boundary, CPROPS extrapolates from the observed boundary (\S\ref{sec_gmas_cprops}) to a sensitivity of 0 K using a weighted, linear least-squares fit  that takes into account the brightness temperature profile within the GMA.
The difference between the GMA luminosity before ($L_{\mathrm{CO,noex}}$)  and after ($L_{\mathrm{CO}}$) the extrapolation is therefore determined by the magnitude of the brightness temperature gradient within the GMA and the CO brightness of the observed boundary. $L_{\mathrm{CO}}$ is typically higher than $L_{\mathrm{CO,noex}}$ by a factor of $\sim$ 3 - 5 for GMAs in the central region, a factor of 3 - 3.5 higher in the bar and spiral arms, and $\sim$ 2.5 times higher in the inter-arm regions.
This is because the brightness temperature gradient within the GMAs varies between the regions \citep{Col14}.
The central region profile is the steepest, and the inter-arm profile is the most shallow. Moreover, the central region is more crowded than other regions, i.e., higher merge levels or observed boundaries, thereby resulting in more emission being abandoned as watershed.

The results of luminosity mass  ($M_{\mathrm{c}}$) are shown in panel (a).
CPROPS  estimates $M_{\mathrm{c}}$ as
\begin{equation}
\begin{split}
M_{\mathrm{c}}[\mathrm{M_{\odot}}]= & \frac{X_{\mathrm{CO}}}{2\times10^{20}[\mathrm{cm^{-2}(K\: km\: s^{-1})^{-1}}]}  \\
                                    & \times 4.4L\mathrm{_{CO}}[\mathrm{K\: km\: s^{-1}\: pc^{2}}],
\end{split}
\end{equation}
where $X_{\mathrm{CO}}$ represents the CO-to-H$_{2}$ conversion factor ($X_{\mathrm{CO}}$). A Galactic value of 2 $\times$ 10$^{20}$ cm$^{-2}$ (K km s$^{-1}$)$^{-1}$ is adopted in this work.    The uncertainty of $X_{\mathrm{CO}}$ is discussed in \S\ref{sec_xco}.  In short, we do not expect large  variations in the value of $X_{\mathrm{CO}}$ between the Milky Way and M100, and among the defined environments.  The typical uncertainty determined via the 
bootstrapping technique is a factor of $\sim$ 2. The median mass of the  GMAs is $\sim$ 8.8 $\times$ 10$^{6}$ M$_{\sun}$. The total mass enclosed within the 165 GMAs is about 2.5 $\times$ 10$^{9}$ M$_{\sun}$.

The CNR GMAs are more massive than the bar, spiral, and inter-arm GMAs in Figure \ref{FIG_Env_Properties}(a). 
The $M_{\mathrm{c}}$ values of all CNR GMAs are greater than $\sim$ 10$^{7}$ M$_{\sun}$. The profile peaks at $\sim$ 10$^{8}$ M$_{\sun}$.  Here, we note that due to the small number of the CNR GMAs, the error bar of each histogram bin is relatively large. Thus, the profile of the distribution of the CNR GMAs should be interpreted with caution. Despite this condition, the $p$-values suggests that the CNR GMAs are significantly different (more massive) from that in any other environments.

The distributions of the bar, spiral, and inter-arm GMAs peak at the same $M_{\mathrm{c}}$ value of $\sim$ 10$^{7}$ M$_{\sun}$, but  the shapes of the profiles are different. 
 The profile of the bar GMAs is relatively symmetric, while the spiral profile shows an increase in relative fraction towards the higher $M_{\mathrm{c}}$  ($\sim$ 2 $\times$ 10$^{7}$ M$_{\sun}$) values.
 In other words, the bar GMAs have relatively lower mass compared to the spiral GMAs.
However, the $p$-value ($p$ $=$ 0.7367) of spiral-bar GMAs suggests that there is no difference between the mass of these two populations. 
On the other hand, the inter-arm GMAs appear to have lower mass compared with the bar GMAs from Figure \ref{FIG_Env_Properties}(a). 
The corresponding p-values suggest that the  typical mass of the inter-arm GMAs is significantly different from those of the bar and spiral GMAs.

Although $M_{\mathrm{c}}$ shows some variations among the environments,  ($R_{\mathrm{c}}$) is similar for all environments, as seen in Figure \ref{FIG_Env_Properties}(b). 
$R_{\mathrm{c}}$ is calculated as
\begin{equation}
R_{\mathrm{c}} = 1.91\sigma _{\mathrm{r}}\mathrm{(0K)}.
\end{equation}
Here, $\sigma_{\mathrm{r}}$(0K) denotes the root-mean-squared spatial size of the  GMA, which is estimated from the geometric mean of the second centralized moment\footnote{The second centralized moment is commonly called the variance and is denoted as $\sigma^{2}$. The rms (standard deviation) $\sigma$ denotes the square root of the variance.} of the intensity distribution along the major and minor axes of the actual cloud boundary. The second moments of the major and minor axes are deconvolved by the beamsize before obtaining  the geometric mean. 
Parameter  $\sigma_{\mathrm{r}}$(0K) is related to $R_{\mathrm{c}}$ by a constant of 1.91 assuming the fixed density profile of a spherical cloud \citep{Sol87}. Eventually, $R_{\mathrm{c}}$ is around 50\% larger than the observed radius (the bright high-density CO region without extrapolation and deconvolution). The typical uncertainty of $R_{\mathrm{c}}$ is $\sim$ 60\%. The profiles of $R_{\mathrm{c}}$ peak at $\sim$ 300 pc for all environments, and so do the median values.  The large $p$-values ($p$ $>$ 0.1) also suggests that the $R_{\mathrm{c}}$ is similar among the environments.

The inter-arm GMAs show the largest relative proportion in the large-valued region of the plot with $R_{\mathrm{c}}$ $\approx$ 700 pc.
Since the $M_{\mathrm{c}}$ values of the inter-arm GMAs lie at the lower end of the mass distribution, this would suggest that these inter-arm GMAs are  more extended for a given mass. The extended inter-arm GMAs are also visible in the left panel of Figure \ref{FIG_Env_Properties}, where many GMAs in the inter-arm regions exhibit a larger size when compared with that of GMAs with comparable mass in the spiral arms.

The mass per unit area can also be quantified.
Figure \ref{FIG_Env_Properties}(c) depicts the distribution profile of the surface density ($\Sigma_{\mathrm{H_{2}}}$) that is computed with the use of $M_{\mathrm{c}}$ and $R_{\mathrm{c}}$ as
\begin{equation}
\Sigma_{\mathrm{H_{2}}} = \frac{M_{\mathrm{c}}}{R_{\mathrm{c}}^{2}\pi }.
\end{equation}
The CNR GMAs show higher $\Sigma_{\mathrm{H_{2}}}$ values when compared with those of other regions. The peak of the profile and the median value are observed in the range of 200 - 300 M$_{\sun}$ pc$^{-2}$, which  is about 10 times higher than that of other environments. The lowest and highest $\Sigma_{\mathrm{H_{2}}}$ values of the CNR GMAs are $\sim$ 50 and 1300 M$_{\sun}$ pc$^{-2}$, respectively.  The $p$-values also imply that the CNR GMAs are different from those in other regions. 
 The profiles of the bar, spiral, and inter-arm GMAs peak at the same value of around 30  M$_{\sun}$ pc$^{-2}$.  Both the profile and $p$-value suggests that bar GMAs resemble spiral GMAs. However, the inter-arm GMAs exhibit a secondary excess at the low-$\Sigma_{\mathrm{H_{2}}}$ end of  10  M$_{\sun}$ pc$^{-2}$. The small $p$-values  suggest that the difference between the inter-arm and bar/spiral GMAs is significant.
We note that the $\Sigma_{\mathrm{H_{2}}}$ values of our GMAs are lower than that  of the Galaxy, presumably due to the small filling factor of CO emission in the extragalactic observations.

With $M_{\mathrm{c}}$ and $R\mathrm{_{c}}$, we can  examine Larson's mass-size relation for the GMAs in M100. \cite{Lar81} determined that the Galactic GMCs are characterized  as regards the scaling relation as:
\begin{equation}
 M_{\mathrm{c}}\propto R\mathrm{_{c}}^{a}.
\label{EQ_larsons}
\end{equation} 
 \cite{Lar81} derived a power-law index of $a$ $\approx$ 2, implying that GMCs have approximately constant $\Sigma_{\mathrm{H_{2}}}$.
 Since that formulation, many studies on Galactic GMCs have reported similar results using observations with improved sensitivity and larger samples. Several studies of extragalactic GMC populations have also arrived at the same conclusion. However, in theory, it has been suggested that the constant $\Sigma_{\mathrm{H_{2}}}$ may be a result of the limited surface brightness sensitivity of observations, and in reality, the $\Sigma_{\mathrm{H_{2}}}$ values of GMCs span at least two orders of magnitude \citep{Keg89,Vaz97,Bal02,Pan16}.

Owing to the high sensitivity of ALMA, as shown in Figure  \ref{FIG_Env_Properties}(c), $\Sigma_{\mathrm{H_{2}}}$ of GMAs differ by about 10 times within individual environments, and more than two orders of magnitude among the environments, allowing us to obtain an insight into the $M_{\mathrm{c}}$-$R\mathrm{_{c}}$ relation  with larger  dynamic range in $\Sigma_{\mathrm{H_{2}}}$.
The relationship between $M_{\mathrm{c}}$ and $R\mathrm{_{c}}$ is presented in Figure \ref{FIG_Env_Scatters}(a). The color and symbols are identical in representation to those in Figure \ref{FIG_Env_Properties}.
The correlation coefficients ($cc$) calculated by  \texttt{corrcoef} of the \texttt{NUMPY} package for each environments are presented in the lower-right corner. 
The parameter $cc$ can be interpreted as: -1.0 to -0.5 or 0.5 to 1.0 for  strong relationship, -0.5 to -0.3 or 0.3 to 0.5 for moderate relationship, -0.3 to -0.1 or 0.1 to 0.3 for weak relationship, and -0.1 to 0.1 for none or very weak relationship. 
The best fit of all GMAs is indicated by the black line, and the slope is shown in the lower-right corner.
The error-weighted fit is performed utilizing the \texttt{POLYFIT} function of \texttt{PYTHON}'s \texttt{NUMPY} package. \texttt{POLYFIT} can be used to fit a polynomial of specified order to data using a least-squares approach. The 1$\sigma$ uncertainty  is estimated from the covariance matrix of the fit. 

  $M_{\mathrm{c}}$ and $R\mathrm{_{c}}$ are strongly correlated with $cc$ $>$ 0.5 except at the CNR.
The  $cc$ value is only 0.07, thus indicating a weak  correlation for the CNR GMAs.
 The best fit of $M_{\mathrm{c}}$ and $R\mathrm{_{c}}$ yields an $a$ value of 1.01 $\pm$ 0.14 with all GMAs. 
This is the combined result of various slopes from $\sim$ 0.3 for the CNR GMAs, to $\sim$ 0.8 for the inter-arm GMAs, to $\sim$ 1.4 for the spiral GMAs, and to $\sim$ 1.6 for the bar GMAs. 
 Given the fact that all environments show a similar $R\mathrm{_{c}}$ range of GMAs, such variation implies a change in the intrinsic properties, such as a physical difference in $M_{\mathrm{c}}$ and gas density (cf. \S\ref{sec_discussion}).

\subsubsection{Velocity Dispersion ($\sigma_{\mathrm{c}}$) and  $R_{\mathrm{c}}$-$\sigma_{\mathrm{c}}$ Relation}
\label{sec_prop_v}

\begin{figure*}
\epsscale{0.95}
\plotone{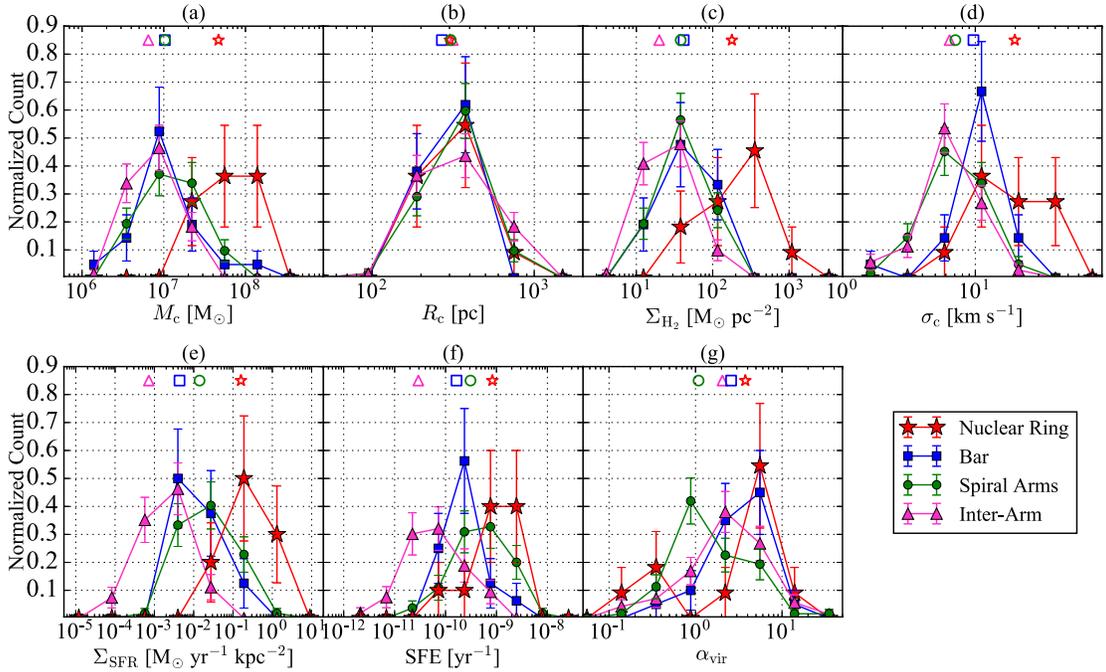}
\caption{Normalized distribution of properties for GMAs in different galactic environments. Filled red stars, blue squares, green circles, and magenta triangles represent GMAs in the CNR, bar, spiral arms, and the inter-arm regions, respectively. Median values of properties are indicated with an open symbol. Error bars represent Poisson errors. In the panels, we plot GMA (a)  mass ($M_{\mathrm{c}}$), (b)  effective radius ($R_{\mathrm{c}}$), (c) surface density ($\Sigma_{\mathrm{H_{2}}}$), (d) velocity dispersion ($\sigma_{\mathrm{c}}$),  (e) star formation rate surface density ($\Sigma_{\mathrm{SFR}}$),  (f) star formation efficiency (SFE), and (g) virial parameter ($\alpha_{\mathrm{vir}}$). \label{FIG_Env_Properties}}
\end{figure*}

Returning to Figure \ref{FIG_Env_Properties}, it is to be noted that  the velocity dispersion ($\sigma_{\mathrm{c}}$) again varies among the environments, as can be observed from panel (d).  $\sigma_{\mathrm{c}}$ is derived from the rms velocity dispersion within the extrapolated (0 K) cloud boundary. Further, $\sigma_{\mathrm{c}}$ is corrected for resolution by subtracting the channel width. The typical uncertainty of $\sigma_{\mathrm{c}}$ is around 70\%. 
The profile peaks at around 10 km s$^{-1}$ for the CNR and bar GMAs. The difference between the CNR and the bar lies in the high-$\sigma_{\mathrm{c}}$ end of the plot ($>$ 20 km s$^{-1}$). 
 The CNR GMAs show   a larger  proportion than the bar GMAs. The corresponding  $p$-values also suggest that the $\sigma_{\mathrm{c}}$ values  of the CNR and bar GMAs are different. The spiral and inter-arm GMAs have lower values of $\sigma_{\mathrm{c}}$, with both peaking at $\sim$ 6 km s$^{-1}$. The $p$-values imply that the GMAs in these two regions are indistinguishable populations with regard to the mean.

For structures at scales considerably larger than a typical GMC, the $\sigma_{\mathrm{c}}$ values that we measure may exhibit broadening due to systematic motions within the galactic disc, such as galaxy rotation. By subtracting a galaxy rotation model from the cloud radial velocities, \cite{Hug13} have shown that these effects have small influence on $\sigma_{\mathrm{c}}$ for structures $<$ 500 pc in CPROPS,  and  this result is insensitive to galaxy types\footnote{\cite{Hug13} performed their analysis on M51 and LMC. The former is a large grand design spiral galaxy, while the latter is a small disrupted barred spiral galaxy. The inclinations of all of the three galaxies (M51, LMC, and M100) are as low as $<$ 30$^{\circ}$. The CO observations covered the entire molecule-rich region in these three galactic discs. Based on these results, we assume that the results of M51 and LMC are applicable to M100.}.  The typical difference between GMA with and without rotation subtraction is  smaller than our velocity resolution of 5 km s$^{-1}$. Therefore, we ignore the effect of galaxy rotation in this work. Moreover, since the difference in $\sigma_{\mathrm{c}}$ before and after galaxy rotation subtraction is significantly smaller than the difference between the CNR GMAs and other regions, the high $\sigma_{\mathrm{c}}$ values  of CNR GMAs are likely true.

\begin{figure*}
\epsscale{1.1}
\plotone{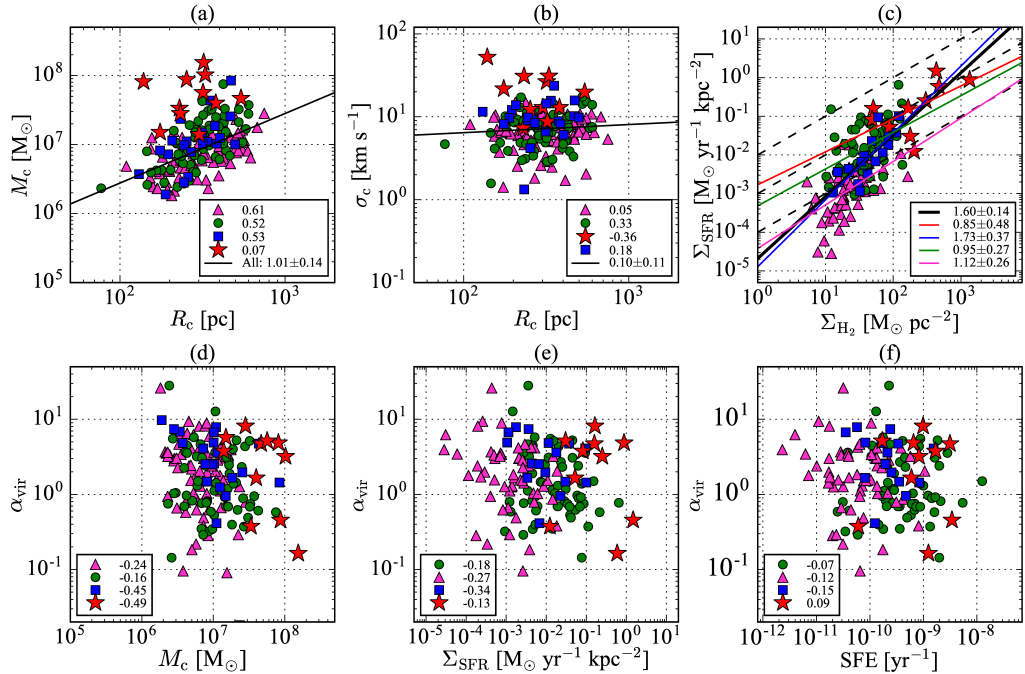}
\caption{Scatter plots of GMAs in different environments. Color and symbols are the same as in Figure \ref{FIG_Env_Properties}. (a) $M_{\mathrm{c}}$ versus $R_{\mathrm{c}}$. Black solid line indicates the best fit of $M_{\mathrm{c}}$-$R_{\mathrm{c}}$ relation with all GMAs. The slope is 1.01 $\pm$ 0.14.   Correlation coefficients of each environment are provided in the lower-left corner of the plots.  (b)  $\sigma_{\mathrm{c}}$ versus $R_{\mathrm{c}}$. The best fit with all populations is shown by a black line. The slope of the best fit is presented in the lower-right corner along with the  correlation coefficients of each environments. 
(c) $\Sigma_{\mathrm{SFR}}$ versus $\Sigma_{\mathrm{H_{2}}}$, the so-called Kennicutt-Schmidt relation. The three dashed lines denote SFE values of 10$^{-10}$ (lower),  10$^{-9}$, and 10$^{-8}$ (upper) yr$^{-1}$, in that order.  The derived slopes of each environments are shown in the  lower-right corner  of the plot.   Panels  (d), (e), and (f) display  $\alpha_{\mathrm{vir}}$ versus $M_{\mathrm{c}}$, $\Sigma_{\mathrm{{SFR}}}$, and SFE, respectively. 
 \label{FIG_Env_Scatters}}
\end{figure*}

Figure \ref{FIG_Env_Scatters}(b) shows the scatter plot of $\sigma_{\mathrm{c}}$ versus $R_{\mathrm{c}}$, that is, the second Larson's relation in the form of
\begin{equation}
 \sigma_{\mathrm{c}}\propto R\mathrm{_{c}}^{b}.
\end{equation}
 The two variables are at most very weakly correlated as implied by the $cc$ value. 
The best fit of all populations yields a power-law index of $b$ $=$ 0.10 $\pm$ 0.11. 
The two variables are moderately correlated in the CNR and spiral arms, and show weak-to-no correlation in the inter-arm and bar regions. In general, there is no obvious $R_{\mathrm{c}}$-$\sigma_{\mathrm{c}}$ relation for GMAs in M100.

\def\arraystretch{1.4}
\begin{table*}
\caption{$p$-value of GMA properties between different environments. For each environment, we calculated the $p$-value with three other  environments separately.    
The $p$-value  is interpreted in the following usual way:
$<$ 0.01 for \emph{highly significant} (two environments are significantly different),  0.01 -- 0.05 for   \emph{significant},  0.05 -- 0.1 for \emph{suggestive}, and $>$ 0.1 for \emph{non-significant} (two environments are likely the same).  For convenience, in Table \ref{TAB_p_value} the $p$-value are marked with $\times$, $\triangle$, $\lozenge$, and $\bigcirc$ for  highly significant difference, significant difference, suggestive of difference, and non-significant difference, respectively.
}
\begin{minipage}{0.45\textwidth}
\begin{tabular}{|c|cccc|}
\hline
$\mathrm{\mathbf{M_{c}}}$   & CNR    & Bar    & Arm                      & ITA    \\
\hline
CNR & 1.0000 & --     & --                       & --     \\
Bar & 0.0068($\times$) & 1.0000 & --                       & --     \\
Arm & 0.0054($\times$) & 0.7367($\bigcirc$) & 1.0000                   & --     \\
ITA & 0.0025($\times$) & 0.0715($\triangle$) & 0.0005($\times$) & 1.0000  \\
\hline
\end{tabular}
\end{minipage}
\hspace{20pt}
\begin{minipage}{0.45\textwidth}
\begin{tabular}{|c|cccc|}
\hline
$\mathrm{\mathbf{R_{c}}}$   & CNR    & Bar    & Arm                      & ITA    \\
\hline
CNR & 1.0000 & --     & --                       & --     \\
Bar & 0.8352($\bigcirc$) & 1.0000 & --                       & --     \\
Arm & 0.4564($\bigcirc$) & 0.4841($\bigcirc$) & 1.0000                   & --     \\
ITA & 0.2360($\bigcirc$) & 0.1985($\bigcirc$) & 0.4319($\bigcirc$) & 1.0000  \\
\hline
\end{tabular}
\end{minipage}
\\
\begin{minipage}{0.45\textwidth}
\begin{tabular}{|c|cccc|}
\hline
$\mathrm{\mathbf{\Sigma_{H_{2}}}}$   & CNR    & Bar    & Arm                      & ITA    \\
\hline
CNR & 1.0000 & --     & --                       & --     \\
Bar & 0.0418($\triangle$) & 1.0000 & --                       & --     \\
Arm & 0.0377($\triangle$) & 0.4614($\bigcirc$) & 1.0000                   & --     \\
ITA & 0.0293($\triangle$) & 0.0130($\triangle$) & 0.0022($\times$) & 1.0000  \\
\hline
\end{tabular}
\end{minipage}
\hspace{20pt}
\begin{minipage}{0.45\textwidth}
\begin{tabular}{|c|cccc|}
\hline
$\mathrm{\mathbf{\sigma_{c}}}$   & CNR    & Bar    & Arm                      & ITA    \\
\hline
CNR & 1.0000 & --     & --                       & --     \\
Bar & 0.0196($\triangle$) & 1.0000 & --                       & --     \\
Arm & 0.0057($\times$) & 0.0197($\triangle$) & 1.0000                   & --     \\
ITA & 0.0042($\times$) & 0.0033($\times$) & 0.2112($\bigcirc$) & 1.0000  \\
\hline
\end{tabular}
\end{minipage}
\\
\begin{minipage}{0.45\textwidth}
\begin{tabular}{|c|cccc|}
\hline
$\mathrm{\mathbf{\alpha_{c}}}$   & CNR    & Bar    & Arm                      & ITA    \\
\hline
CNR & 1.0000 & --     & --                       & --     \\
Bar & 0.8242($\bigcirc$) & 1.0000 & --                       & --     \\
Arm & 0.2437($\bigcirc$) & 0.0938($\lozenge$) & 1.0000                   & --     \\
ITA & 0.5473($\bigcirc$) & 0.3023($\bigcirc$) & 0.3845($\bigcirc$) & 1.0000  \\
\hline
\end{tabular}
\end{minipage}
\hspace{20pt}
\begin{minipage}{0.45\textwidth}
\begin{tabular}{|c|cccc|}
\hline
$\mathrm{\mathbf{\Sigma_{SFR}}}$   & CNR    & Bar    & Arm                      & ITA    \\
\hline
CNR & 1.0000 & --     & --                       & --     \\
Bar & 0.0471($\triangle$) & 1.0000 & --                       & --     \\
Arm & 0.0622($\lozenge$) & 0.0787($\lozenge$) & 1.0000                   & --     \\
ITA & 0.0395($\triangle$) & 0.0323($\triangle$) & 0.0015($\times$) & 1.0000  \\
\hline
\end{tabular}
\end{minipage}
\\
\begin{minipage}{0.45\textwidth}
\begin{tabular}{|c|cccc|}
\hline
SFE   & CNR    & Bar    & Arm                      & ITA    \\
\hline
CNR & 1.0000 & --     & --                       & --     \\
Bar & 0.0298($\triangle$) & 1.0000 & --                       & --     \\
Arm & 0.1890($\bigcirc$) & 0.0104($\triangle$) & 1.0000                   & --     \\
ITA & 0.0117($\triangle$) & 0.0597($\lozenge$) & $<$ 0.0001($\times$) & 1.0000  \\
\hline
\end{tabular}
\end{minipage}
\hspace{20pt}
\label{TAB_p_value}
\end{table*}

\subsection{Star Formation Activity of GMAs in Galactic Environments}
\label{sec_sf}

Figure \ref{FIG_Env_Properties}(e) displays the distribution of the star formation rate surface density ($\Sigma_{\mathrm{SFR}}$). $\Sigma_{\mathrm{SFR}}$ represents  the SFR per unit area as defined by 
\begin{equation}
\Sigma_{\mathrm{SFR}}[\mathrm{M_{\odot}\: yr^{-1}\: kpc^{-2}}]=10^{6}\frac{\mathrm{SFR(H\alpha)\: [M_{\odot}\: yr^{-1}]}}{\pi R_{\mathrm{c}}^{2}\: \mathrm{[pc^{2}]}}.
\end{equation}
 The uncertainty in $\Sigma_{\mathrm{SFR}}$ is a factor of a few as estimated from the error propagation.
The $\Sigma_{\mathrm{SFR}}$ value appears to be highest in the CNR,   followed by the spiral, bar, and inter-arm regions. The profile of the CNR peaks at 0.2 M$_{\sun}$ yr$^{-1}$ kpc$^{-2}$, which is consistent with the radial averaged $\Sigma_{\mathrm{SFR}}$ of this area as derived by \cite{Won02}.  The spiral GMAs show a peak at $\sim$ 0.02 M$_{\sun}$ yr$^{-1}$ kpc$^{-2}$, but a significant fraction is also observed between 10$^{-3}$ and 0.01 M$_{\sun}$ yr$^{-1}$. The bar GMAs span a similar range as that of the spiral GMAs, but the peak shifts to $\sim$ 10 times lower. The inter-arm GMAs peak at the same $\Sigma_{\mathrm{SFR}}$ as that of the bar GMAs, but there is a significant fraction of low-$\Sigma_{\mathrm{SFR}}$ objects with $\Sigma_{\mathrm{SFR}}$ $<$ 10$^{-3}$ M$_{\sun}$ yr$^{-1}$ kpc$^{-2}$. The $p$-values in Table \ref{TAB_p_value} imply suggestive-to-significant levels of difference between the environments.

The final plot in Figure \ref{FIG_Env_Properties} shows the distribution of the star formation efficiency  (SFE).  SFE describes the number of stars formed per year in a GMA, and is formulated as
\begin{equation}
\mathrm{SFE\: [yr^{-1}]}=\frac{\mathrm{SFR(H\alpha)\: [M_{\odot}\: yr^{-1}]}}{M_{\mathrm{c}}\: \mathrm{[M_{\odot}]}}.
\end{equation}
Most of the CNR GMAs show a high SFE $\geqslant$ 5 $\times$ 10$^{-10}$  yr$^{-1}$, and the profile peaks at $\sim$ (1 -- 2) $\times$ 10$^{-9}$  yr$^{-1}$. This SFE is consistent with the values averaged over the nearby CNR regions in the study by \cite{Ken98}. 
For the three environments in the galactic disc, the peak of SFR decreases from 8 $\times$ 10$^{-10}$  yr$^{-1}$ of spiral GMAs, to 2 $\times$ 10$^{-10}$  yr$^{-1}$ of bar GMAs, and to 7 $\times$ 10$^{-11}$  yr$^{-1}$ of inter-arm GMAs. These differences are likely to be true, as suggested by the $p$-values. 
As $M_{\mathrm{c}}$  requires an assumption of $X_{\mathrm{CO}}$  to compute SFE, i.e., $\propto \mathrm{SFR}/(L_{\mathrm{CO}}X_{\mathrm{CO}})$,  SFE would decrease if a higher $X_{\mathrm{CO}}$ is adopted, and vice versa.
Here, we note that we would not expect SFE to vary by a factor of more than a few units among the environments given that $X_{\mathrm{CO}}$   is relatively constant across the disc of M100 (cf. \S\ref{sec_xco}).

\section{Discussion and Comparison With Previous Studies}
\label{sec_discussion}

Information linking molecular gas and star formation to galactic-scale environments is mainly available at sub-kpc resolution. 
However, cloud observations must be made on scales of tens of pcs to fully resolve their properties. Given that the typical separation of Galactic GMCs is a few 100 pc to kpc \citep{Sol87,Kod06}, in this work, we have assumed that the GMA properties  can generally represent the underlying local GMC properties. This section compares GMA properties with GMC properties in literature.

\subsection{Massive GMAs at the Circumnuclear Region}
\label{sec_cnr_gmas}

The ambient pressure in the ISM  surrounding the GMCs plays a role in regulating their $\Sigma_{\mathrm{H_{2}}}$ and $\sigma_{\mathrm{c}}$.  
The tendency of CNR GMAs to be  more massive  with higher $\Sigma_{\mathrm{H_{2}}}$ and $\sigma_{\mathrm{c}}$  than those in the disc environments has been already suggested through several previous observational studies with tens-of-pc resolutions \citep[e.g.,][]{Hey01,Oka01,Ros05,Fie11,Hug13}.
The intercloud pressure required to confine  GMCs lies in the range from $\sim$ 10$^{4}$ to 10$^{6}$ K cm$^{-3}$. The origin of the external (ISM) pressure that is required to explain such a wide range of pressure is not clear thus far. The possible sources are the thermal and magnetic pressures of the ambient ISM, the weight of gas, stars, and dark matter \citep{Elm89}, recoil pressure from the release of H atoms from GMCs by UV radiation \citep{Fie09}, and pressure from large-scale turbulence \citep{Hei09}. Though these sources are likely common in extreme environments such as CNRs, future studies are required to elucidate the nature of the pressure.

\subsection{The Bar GMAs}
\label{sec_disc_gmas}
 The bar regions in galaxies offers an ideal laboratory to study the interplay between the kpc-scale dynamics (gas flows, shear) and the pc-scale GMCs. 
The bar GMAs of M100 have relatively low $M_{\mathrm{c}}$ values compared with those of other gas-rich environments, and the bar SFE also lies in the lower half of the SFE spread in the spiral. 
Several mechanisms have been proposed to explain small $M_{\mathrm{c}}$ and low SFE of bar clouds.

It is known that the orbital motions within the bar induce intense shear (velocity gradient)  in the ISM, which could prevent the formation of massive objects \citep{Ath92,Hop12}.  The shear  at the bar also boosts the $\sigma_{\mathrm{c}}$ value at the bar. We can compare the observed shear and the required shear to destroy GMCs.
The de-projected velocity difference across the high-density CO bar ridges is $\sim$154 km s$^{-1}$, measured from the velocity field\footnote{A clip at the 5$\sigma$ level was used to make the first moment map from the spectral data cube.}  The width of the bar is 400 pc,  and thus, the shear (velocity gradient) is about  0.4 km s$^{-1}$ pc$^{-1}$. Further, the typical brightness temperature is $\sim$ 2 K along the bar. Thus,  the beam filling factor at the bar is around 0.2, given that the typical GMC temperature  is $\sim$ 10 K.
The unresolved bar GMAs have typical  $\Sigma_{\mathrm{H_{2}}}$ values  of  30 M$_{\sun}$ pc$^{-2}$.
Therefore, the intrinsic $\Sigma_{\mathrm{H_{2}}}$ of molecular clouds is 30/0.2 $\approx$ 150 M$_{\sun}$ pc$^{-2}$. Assuming that the cloud radius is $R_{\mathrm{c}}$ $\approx$ 10 pc, the escape velocity from the cloud is $\sqrt{2\pi G\Sigma _{\mathrm{H_{2}}}R_{\mathrm{c}}}\approx6$ km s$^{-1}$  \citep{Kod06b}. Thus, a shear  of 6/10 $=$ 0.6  km s$^{-1}$ pc$^{-1}$ is necessary to destroy the molecular cloud. The  observed  velocity gradient is smaller than this value, but the difference is not large. Thus, the influence of shear on the molecular clouds cannot be ruled out.
      
Tubbs (1982) suggested that GMCs entering the dust lane could be dispersed due to their high velocities ($>$ 20 -- 60 km s$^{-1}$) relative to the  dust lane gas.  
A higher velocity threshold of 80 -- 170 km s$^{-1}$ was subsequently suggested by \cite{Rey98}.
In the intermediate radial range of the bar ($\sim$ 2 kpc) where the H$\alpha$ image clearly shows that the star formation is weak,  the rotation velocity ($v_{r}$) of the gas is 217 km s$^{-1}$ \citep{Sof99}. The  rotation velocity  of the density wave  ($\Omega_{p}r$) at this radius is 70 km s$^{-1}$, given that the pattern speed ($\Omega_{p}$) of M100 is 35 km s$^{-1}$ kpc$^{-1}$ \citep{She02}.  A GMA thus enters the density wave with a velocity as high as 147 km s$^{-1}$. Therefore, the entry velocity is sufficiently high to disrupt GMAs in this scenario.
On the other hand, at the bar end  (at $\sim$ 4.5 kpc with $v_{r}$ $\approx$ 238 km s$^{-1}$) where more star formation occurs, the entry velocity is almost two times lower, i.e.,  $\sim$ 80 km s$^{-1}$.
	
 High-resolution simulations  (1.5 pc)  predict that the frequent cloud interactions in the bar not only generate tidal features in which a large amount of small and low $\Sigma_{\mathrm{H_{2}}}$ is formed as suggested by observations, but also cause mergers to build up massive GMAs. These massive GMAs are very likely to be resolved, or at least be fully isolated from  the surrounding gas in our observations. We found that the bar GMAs do correspond to a higher proportion of high-mass GMAs when compared with other disc regions. However, the number is small, and further investigation with a large sample of barred galaxies is required to confirm their existence.
In addition to the high frequency, clouds formed in the bar region typically collide faster than those in the spiral \citep{Fuj14b}. The unproductive collisions in the bar region lower the SFE even though collisions are more common in the bar region. 

\subsection{The Spiral and Inter-arm GMAs}
\label{sec_spita}
The dynamics of spiral arms is believed to regulate the cycling of molecular gas in the galactic disc. 
 Inter-arm GMCs are remnants of massive GMAs that were previously in the spiral arms as a result of shearing forces that  tear apart  the inter-arm GMAs.   
Consequently, the inter-arm GMAs appear more diffuse (with larger $R_{\mathrm{c}}$ values  for a given $M_{\mathrm{c}}$, i.e., lower $\Sigma_{\mathrm{H_{2}}}$ and shallower $M_{\mathrm{c}}$-$R_{\mathrm{c}}$ relation) and with lower $\alpha_{\mathrm{vir}}$  (cf. \S\ref{sec_dynamical_gmas}).
Stellar feedback such as photodissociation by massive star and supernova explosions should also contribute to cloud disruption since star formation occurs in the inter-arm regions as well (see Figure \ref{FIG_Halpha}). However, stellar feedback may not be the dominant (effective) mechanism for GMA disruption given their masses \citep{Kod09,Mei13}. 
The inter-arm GMCs are predicted to subsequently coagulate into massive GMAs due to spiral arm streaming motions during the next spiral arm passage \citep{Wil97,Kod09,Dob12,Dob13,Mei13,Mei15}.
The mass distribution of our GMAs also qualitatively supports this scenario. 

\subsection{Kennicutt-Schmidt Relation}
\label{sec_ks}
Variation in SFE is also revealed in the relationship between $\Sigma_{\mathrm{H_{2}}}$ and $\Sigma_{\mathrm{SFR}}$, which is the so-called Kennicutt-Schmidt (K-S) relation (Figure \ref{FIG_Env_Scatters}(c)). The three dashed lines denote SFE values of 10$^{-10}$ (lower),  10$^{-9}$, and 10$^{-8}$ (upper) yr$^{-1}$, in that order. The black line represents the best fit of the relation
\begin{equation}
\Sigma _{\mathrm{SFR}}\propto \Sigma _{\mathrm{H_{2}}}^{N}
\label{eq_ks}
\end{equation}
using all GMAs.
An index of $N$ $\approx$ 1.5 is subsequently derived. 
The value is consistent with the  disc-averaged relation suggested by the original study of \cite{Ken98}, and the result of \cite{Ken98} is interpreted as that galaxies are governed by a universal star formation law.
However, it is seen from the figure that the environments in question correspond to different regimes on the $\Sigma_{\mathrm{H_{2}}}$-$\Sigma_{\mathrm{SFR}}$ plane, and that  putting the universal \emph{law} into doubt.
For this reason, we also fit the GMAs for individual environments using Equation (\ref{eq_ks}). The power-law index $N$ decreases from a superlinear relation ($\sim$ 1.7) for the bar GMAs to an almost linear relation ($\sim$ 1.0) for the spiral and inter-arm GMAs, and to a sublinear relation ($\sim$ 0.8) for the CNR GMAs,  indicating a non-universal K-S relation.

The small $N$ value  of the CNR GMAs is due to the large scatter in $\Sigma_{\mathrm{H_{2}}}$, $\Sigma_{\mathrm{SFR}}$, and SFE values, as seen in Figure \ref{FIG_Env_Properties}. It is to be noted that the defined CNR region contains the ring-like CNR itself, the nuclear bar, and the galactic center. CNR GMAs may display a variety of behaviors as their properties are regulated by different environments or a mixture of them. 

A slope of 1.7 may imply that a combination of multiple mechanisms is necessary for forming stars in the bar.
The frequent collisions in the bar can trigger some star formation that would drive the slope close to  2.0 \citep{Tan00}. On the other hand,   if star formation is induced by large-scale gravitational instabilities, a slope of $N$ $\approx$ 1.5 is expected. 
A slope of 1.7 might be the result of these two star-forming mechanisms.
 
However, the star formation mechanism in the spiral regions is not as straightforward as implied by the K-S relation. 
The derived $N$ suggests that the spiral GMAs have fixed SFE, and therefore $N$ $\approx$ 1. At face value, a slope of unity would imply that the SFE is determined by the intrinsic properties of a cloud and not strongly correlated with the environment. This is doubtful since $\Sigma_\mathrm{SFR}$ of the spiral GMAs is more scattered than that for the bar while their $\Sigma_{\mathrm{H_{2}}}$ values are comparable. The GMA properties and their relation with the star formation mechanism maybe more complicated than expected in this region.

The K-S relation of the inter-arm GMAs increases steeply at low $\Sigma_{\mathrm{H_{2}}}$ (5 -- 20 M$_{\sun}$ pc$^{-2}$), and this trend is not observed in any other environment. In this narrow range, $\Sigma_{\mathrm{SFR}}$ scatters by $\sim$ 100 times. This result indicates a change in the star formation process and gas properties at $\sim$ 10 M$_{\sun}$ pc$^{-2}$.
The steep slope at $\sim$ 10 M$_{\sun}$ pc$^{-2}$  is often interpreted as the surface density at which gas becomes molecular in simulations \citep{Kru09,Dob09}. 
Below this threshold, the gas mainly comprises low-density and unbound gas components and shows a weak correlation with the SFR. 
The phase change scenario would imply that the inter-arm GMAs contain larger fractions of the atomic gas component than in other regions. There are no high-resolution HI data of M100 that allow us to distinguish the arm-to-inter-arm variation; these data are not available for most of the nearby galaxies also. Nonetheless, gas phase change between the arm and inter-arm regions has been observed in the Milky Way \citep{Kod16}. Interstellar gas becomes molecular as it enters spiral arms, but is dissociated back to the atomic phase upon leaving the arms. Such a phase change occurs across the entire Galactic disc.

\subsection{Dynamical State of the GMAs}
\label{sec_dynamical_gmas}
\subsubsection{Virial Parameter}
We find little evidence for a correlation between $R_{\mathrm{c}}$ and $\sigma_{\mathrm{c}}$. This is in conflict with the classic relation derived for the Galactic GMCs by \cite{Lar81} and \cite{Sol87}, who found a power-law index of 0.38 and 0.50, respectively. The Galactic $R_{\mathrm{c}}$-$\sigma_{\mathrm{c}}$ relation was interpreted as evidence that GMCs are supported by internal turbulence, and are gravitationally bound.
On the contrary, our result would indicate that the GMAs are in diverse dynamical states.
In fact, a lack of correlation between these two variables has also been reported in recent extragalactic observations with $\leq$ 50 pc resolution for a variety of galaxies  \citep[e.g., LMC, M33, M51, M101, NGC 628, and NGC 6946;][]{Hug13,Col14,Reb15}.

We next  quantify the dynamical state of the GMAs by calculating their  $\alpha_{\mathrm{vir}}$ (virial parameter). $\alpha_{\mathrm{vir}}$ is a measure of gravitational binding, and it is defined as the ratio of the GMA virial mass ($M_{\mathrm{vir}}$) to  $M_{\mathrm{c}}$  as
\begin{equation}
\alpha _{\mathrm{vir}}=\frac{M_{\mathrm{vir}}}{M_{\mathrm{c}}}\approx \frac{1040R_{\mathrm{c}}\sigma \mathrm{_{c}}^{2}}{M_{\mathrm{c}}}.
\label{EQ_alpha}
\end{equation}
An  $\alpha_{\mathrm{vir}}$ value of $<$ 2 indicates that the GMA is gravitationally bound and vice versa.

$\alpha_{\mathrm{vir}}$  spans a wide range as shown in Figure  \ref{FIG_Env_Properties}(g). The peak values of the profiles decrease from $\sim$ 5 for the CNR and bar GMAs, to $\sim$ 2 for the inter-arm GMAs, and to $\sim$ 0.9 for the spiral GMAs. The same trend is also observed for the median values, but it is less distinguishable. This suggests that the CNR, bar, and the inter-arm GMAs are either not bound or only marginally bound, while the spiral GMAs are in general self-gravitating.  However, the $p$-values suggest that the difference in the means of the environments is not statistically significant. Nonetheless, Figure  \ref{FIG_Env_Properties}(g)  still implies that the dynamical states of GMAs are truly diverse as $\alpha_{\mathrm{vir}}$ spans nearly two orders of magnitude.

In Figure \ref{FIG_Env_Scatters}(d), we plot $\alpha_{\mathrm{vir}}$ of the GMAs as a function of $M_{\mathrm{c}}$. The correlation coefficients of  0.15 $<$ $\left|cc\right|$ $<$ 0.50    suggest a weak-to-moderate anticorrelation between $\alpha_{\mathrm{vir}}$ and $M_{\mathrm{c}}$. This dependence suggests that overall the high-$M_{\mathrm{c}}$ GMAs in M100 tend to be more bound than low-$M_{\mathrm{c}}$ GMAs. The dependence is also seen in the high-resolution observation of GMCs of M51 \citep{Col14} and simulations \citep{She10,Fuj14}. 

The formation of stars is caused by the fragmentation of collapsing molecular clouds. For a cloud to undergo collapse, the gravitational energy of a cloud has to overcome the kinetic energy that supports it. In this case, $\alpha_{\mathrm{vir}}$ corresponds to $\leqslant$ 2.
Despite the fact that the scale of our GMA is considerably larger than star forming core/clump that directly connects to the star by gravitational collapse,  $\alpha_{\mathrm{vir}}$ of GMAs has a correlation with SFE in the GMAs.
 Figures \ref{FIG_Env_Scatters}(e) and (f) present the correlations of $\alpha_{\mathrm{vir}}$ versus $\Sigma_{\mathrm{SFR}}$ and SFE, respectively. Both $\Sigma_{\mathrm{SFR}}$ and SFE show a possible anticorrelation ($<$ $\|cc\|$ $\approx$ 0.1) with $\alpha_{\mathrm{vir}}$, thereby  indicating that the degree of star formation  may decrease with increasing $\alpha_{\mathrm{vir}}$ as expected. 

For the individual environments, a correlation is observed between GMA properties, $\alpha_{\mathrm{vir}}$, and SFE, particularly in the disc environments.
In spite of the large $\alpha_{\mathrm{vir}}$,  the CNR GMAs are massive and compact (high $\Sigma_{\mathrm{H_{2}}}$) with high SFE.  The spiral GMAs are similar to the CNR GMAs, with high $M_{\mathrm{c}}$ and high SFE,   because the GMAs can collapse (small $\alpha_{\mathrm{vir}}$).
The bar GMAs have similar $\Sigma_{\mathrm{H_{2}}}$ values as those of the spiral GMAs, but the former are less massive GMAs with high $\sigma_{\mathrm{c}}$, presumably due to the high shear and strong shock in the bar. 
Thus, the bar GMAs tend to be unvirialized, and star formation is suppressed. 
The inter-arm GMAs are generally similar to the bar GMAs.  The enhanced shear of the inter-arm region leads to the formation of  low-$M_{\mathrm{c}}$, low-$\Sigma_{\mathrm{H_{2}}}$, and high-$\alpha_{\mathrm{vir}}$, unvirialized GMAs that explain the low SFE in this region.

\subsubsection{Comments on Observational Bias on $\alpha_{\mathrm{vir}}$}

Increasing attention has been devoted to the questions of whether the physical properties of molecular gas measured by spectral line observations represent the intrinsic structures in three dimensions and whether they are sufficient to examine the universality of GMC properties.
Observationally, \cite{Hug13} found that the measured GMC properties and scaling relations depended on instrumental resolution, observational sensitivity, and the choice of decomposition approach \citep[see also][for synthetic observations using simulated galaxies]{Pan15b,Pan16}.
Although we have adopted a commonly used decomposition approach, and the results from our GMAs are generally in agreement with those of high-resolution observations ($<$ 100 pc), we remark that caution should be exercised when comparing the results from different galaxies, instruments, and cloud identification algorithms. The major reason for this ambiguity is the absence of a practical definition of a molecular cloud.

Among the measured properties, $\alpha_{\mathrm{vir}}$ is the most difficult to determine. By applying synthetic observations to a 3D simulated galaxy, we found that it is difficult to determine if a structure is truly gravitationally bound even with observed resolutions as high as 1.5 pc \citep{Pan15b,Pan16}.
This is because small variations in $M_{\mathrm{c}}$, $R_{\mathrm{c}}$, and $\sigma_{\mathrm{c}}$ lead to significant differences in $\alpha_{\mathrm{vir}}$. A large discrepancy in $\alpha_{\mathrm{vir}}$ can lead to inaccurate interpretations of the dynamical state of GMCs, and therefore their potential for star formation. We emphasize that $\alpha_{\mathrm{vir}}$ should be interpreted with caution when considering the dynamical state and environmental dependence of GMCs/GMAs.

\subsection{CO-to-H$_{2}$ Conversion Factor}
\label{sec_xco}
A potential uncertainty as regards $M_{\mathrm{c}}$ and its relevant properties (e.g., SFE) is the constant $X_{\mathrm{CO}}$.
However, we do not expect large variations in the value of $X_{\mathrm{CO}}$ among the environments. 
\cite{San13} solved for spatially-resolved $X_{\mathrm{CO}}$ for M100 and a set of nearby galaxies by assuming that the dust-to-gas ratio is approximately constant on the kpc scale. They found no significant difference in $X_{\mathrm{CO}}$ between nearby galaxies and the Milky Way. Moreover, the $X_{\mathrm{CO}}$ profile is generally flat as a function of the galactocentric radius, except for the galaxy center.  The  $X_{\mathrm{CO}}$ value of galactic centers  is about twice lower  than that of the disc as a result of external pressure or changes in the properties of molecular gas. The $X_{\mathrm{CO}}$ difference between the disc and the galactic center is too small compared to the difference in $M_{\mathrm{c}}$ derived from a constant $X_{\mathrm{CO}}$ in this work. Therefore, instead of adopting a variable $X_{\mathrm{CO}}$ for correcting a relatively small difference, a constant $X_{\mathrm{CO}}$ is adopted in this work. Nonetheless, we note that the SFE of the CNR GMAs may be underestimated due to the possible change in $X_{\mathrm{CO}}$.

\section{Effect of Removing ACA Data on GMA Properties}
\label{sec_obs_effect}

Extragalactic data do not always contain single-dish data to recover the  extended emission and the total flux. For this reason, we compare the GMA properties obtained from the 12-m observations alone and those obtained from the feathered data (the default data set for this work).

\begin{figure}
\epsscale{1}
\plotone{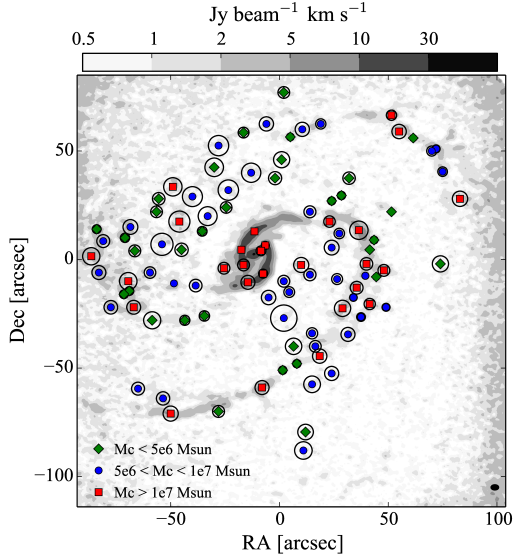}
\caption{Spatial distribution of GMA  $M_{\mathrm{c}}$ and observed radius defined in the 12-m observations.   Green diamonds, blue circles and red squares in the figure denote GMA with $M_{\mathrm{c}}$  $<$ 5 $\times$ 10$^{6}$ M$_{\sun}$,  5 $\times$ 10$^{6}$ $<$ $M_{\mathrm{c}}$  $<$  10$^{7}$ M$_{\sun}$, and $M_{\mathrm{c}}$ $>$  10$^{7}$ M$_{\sun}$, respectively. The circles indicate the \emph{average} radius of the major and minor axes of the observed GMA boundary.The gray scale shows the integrated intensity map of the 12-m observations.
\label{FIG_Mom0_12m}}
\end{figure}

The GMA identification procedures of the 12-m data are the same as that described in \S\ref{sec_data}. We identify a total of 100 GMAs and the total mass of the  GMAs is $\sim$ 1.1 $\times$ 10$^{9}$ M$_{\sun}$, around two times less than that found in the feathered data due to the missing flux. Figure \ref{FIG_Mom0_12m} shows the GMA identified from the data cube of the 12-m observation. The gray scale represents the integrated intensity map of the 12-m observation. The symbols and color of GMAs are the same as those in the left panel of Figure \ref{FIG_Mom0_GMAs}.
As can be observed in the figure, most of the inter-arm GMAs are lost in the 12-m observations due to the missing  extended structures. 
 
\begin{figure}
\epsscale{1.1}
\plotone{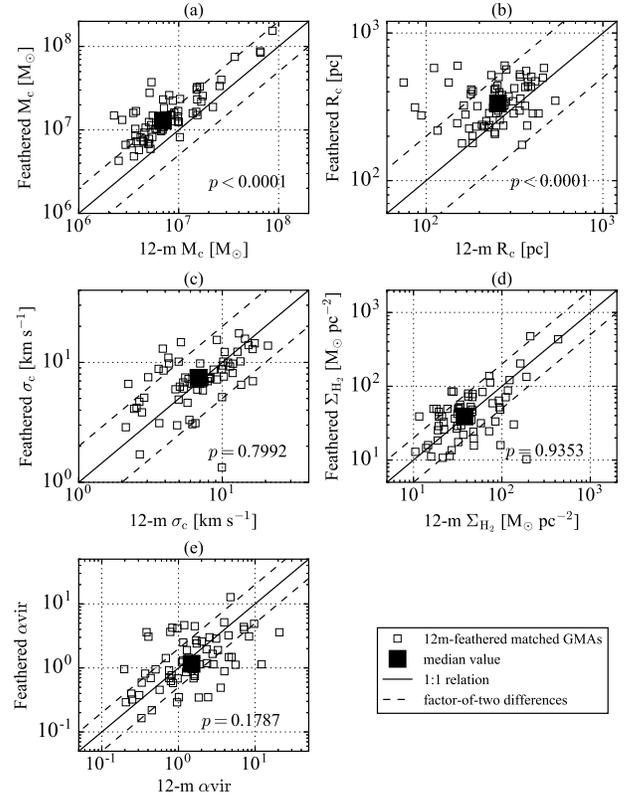}
\caption{One-to-one correlation of  properties of the matched GMA in the 12-m+ACA feathered and 12-m data. Panels  (a) -- (e) present the correlations for $M_{\mathrm{c}}$, $R_{\mathrm{c}}$, $\sigma_{\mathrm{c}}$, $\Sigma_{\mathrm{H_{2}}}$, and $\alpha_{\mathrm{vir}}$, respectively.
The median values of each property are marked with solid squares. Solid and dashed lines mark the 1:1 correlation and factor-of-two differences, respectively. $p$-values of the 12-m and feathered properties are presented in the lower-right corner,
\label{FIG_Prop_12m}}
\end{figure}

Some variation in the GMA $M_{\mathrm{c}}$ and  $R_{\mathrm{c}}$  is observed between the feathered and 12-m GMAs. We identify 70 matched GMAs based on their position and velocity in the two data cubes. One-to-one relations of  the properties are shown in Figure \ref{FIG_Prop_12m} with open squares. The median values of each property are marked with solid squares. Solid and dashed lines mark the 1:1 correlation and factor-of-two differences, respectively.
Feathered data show higher $M_{\mathrm{c}}$ and larger $R_{\mathrm{c}}$ values  because the feathered data capture more  extended emission of a GMA. Further, the $p$-values ($<$ 0.0001) shown in the lower right corner of the plots also suggest that the differences are highly significant. 
On the other hand, removing the ACA data has little effect  on the median $\sigma_{\mathrm{c}}$, $\Sigma_{\mathrm{H_{2}}}$, and $\alpha_{\mathrm{vir}}$. A majority of the GMAs  scatter within a factor of 2. The similarity between the feathered and 12-m GMAs is also reflected in the $p$-values.
The $p$-values of these quantities are as high as  $>$ 0.1, thus indicating that the properties estimated from these two datasets are nearly indistinguishable. \emph{We must nevertheless note that  $M_{\mathrm{c}}$ and  $R_{\mathrm{c}}$ are  used to calculate $\Sigma_{\mathrm{H_{2}}}$ and $\alpha_{\mathrm{vir}}$. While $M_{\mathrm{c}}$ and  $R_{\mathrm{c}}$ are sensitive to the data type,  the  similarity in  $\Sigma_{\mathrm{H_{2}}}$ and $\alpha_{\mathrm{vir}}$ should still be interpreted conservatively.}

It is worth noting that the GMA properties  become increasingly uniform in the 12-m observation, as suggested by the standard deviation of the GMA properties.
The standard deviation of the 12-m GMA properties is 20 -- 30 \% smaller than that of the feathered GMAs because the observation tends to see the compact regions of GMAs.
Such uniformity in the properties may obscure the environmental dependence of the GMA properties. 
In particular,  it could be part of the reason that earlier extragalactic studies concluded that extragalactic GMCs share properties similar with their Galactic counterparts, whereas contradictory results have recently been obtained with  advances in single-dish combination techniques and robustness and reliability of interferometric data.
Therefore, for the study of environmental dependence, the missing flux and extended emission cannot be neglected.

\section{Summary}
\label{Sec_summary}
In this work, we studied the physical properties of giant molecular cloud associations (GMAs) in M100 (NGC 4321)  using the ALMA Science Verification feathered (12-m $+$ ACA) data in $^{12}$CO (1-0) (\S\ref{sec_data}). 
 The spatial resolution of the map was 3.87$\arcsec$ $\times$ 2.53$\arcsec$  ($\sim$ 267
$\times$ 174 pc).
The final image covered  an area of $\sim$ 200$\arcsec$ $\times$ 200$\arcsec$ (14 $\times$ 14 kpc). 
Here, we remark that the low inclination of M100 affords a perfect perspective of galactic structures, allowing us to compare the GMA properties between various environments.

 In the study, 165 reliable GMAs were identified with the use of the cloud finding algorithm CPROPS (\S\ref{Sec_Obj}).  The numbers of GMAs classified according to each environment   are: 11 CNR GMAs, 21 bar GMAs, 62 spiral GMAs, and 71 inter-arm GMAs (\S\ref{sec_classify}). We compared the GMA properties among the galactic environments. The main results are as follows.
 
\begin{enumerate}
\item The CNR GMAs are more massive ($\sim$ 10$^{8}$ M$_{\sun}$) than the bar, spiral, and inter-arm GMAs. The bar GMAs span a similar mass range as that of the spiral GMAs,  but the peak lies on the low-$M_{\mathrm{c}}$ side of the spiral GMAs ($\sim$ 10$^{7}$ M$_{\sun}$). Meanwhile,  the inter-arm GMAs show an increase in relative fraction towards very low-mass end of $\sim$ 4 $\times$ 10$^{6}$ M$_{\sun}$ (\S\ref{sec_prop_MRSig}).

\item  Although the profiles of mass differ between the environments, the range and median values of the GMA effective radius ($R_{\mathrm{c}}$) are similar for all environments. Nonetheless, we found that the inter-arm GMAs are further extended for a given mass compared to GMAs in other environments, i.e., lower molecular mass surface density   ($\Sigma_{\mathrm{H_{2}}}$). In contrast to the inter-arm GMAs, the CNR GMAs are more compact with high $\Sigma_{\mathrm{H_{2}}}$, presumably due to the high-pressure environment (\S\ref{sec_prop_MRSig} and \ref{sec_cnr_gmas}).

\item $M_{\mathrm{c}}$ and $R_{\mathrm{c}}$ are strongly correlated with a correlation coefficient greater than 0.5 for all environments except the CNR. The derived power-law index for the $M_{\mathrm{c}}\propto R\mathrm{_{c}}^{a}$ relation is $a$ $\approx$ 1.0. This is the combined result of various slopes in different environments, implying a change in the intrinsic properties of GMAs (\S\ref{sec_prop_MRSig}).

\item  The profile of velocity dispersion ($\sigma_{\mathrm{c}}$) peaks   at around 10 km s$^{-1}$ for the CNR and bar GMAs, followed by the  spiral arms ($\leqslant$ 10 km s$^{-1}$), and the inter-arm regions ($\sim$ 6 km s$^{-1}$).  
The high $\sigma_{\mathrm{c}}$ of the bar GMAs is  a result of high shear and strong shock in the bar. The shear and shock on the ISM  could also prevent the formation of massive objects, resulting in the relatively low $M_{\mathrm{c}}$ compared with that of the spiral arms (\S\ref{sec_disc_gmas}).

\item We found little evidence for a correlation between $\sigma_{\mathrm{c}}$ and $R_{\mathrm{c}}$ in M100, indicating that the GMAs are in diverse dynamical states. This is indeed observed in terms of the virial parameter ($\alpha_{\mathrm{vir}}$), which spans nearly two orders of magnitude. We found that only the spiral GMAs are in general self-gravitating. In general, the high-$M_{\mathrm{c}}$ GMAs in M100 tend to be more bound than low-$M_{\mathrm{c}}$ GMAs (\S\ref{sec_prop_v} and \ref{sec_dynamical_gmas}).

\end{enumerate}

The sub-kpc resolution offers an ideal opportunity to link molecular gas to star formation because the time-averaged quantities such as SFR require measurements over larger scales to sample the full stellar evolution of individual regions. We calculated the SFR of each GMA by determining the sum of the SFRs of the HII regions within the GMA (\S\ref{sec_calc_sfr}). The main findings are summarized as follows.

\begin{enumerate}
\item 	The star formation rate surface density ($\Sigma_{\mathrm{SFR}}$) and efficiency (SFE)  appear to be highest in the CNR, followed by the spiral, bar, and inter-arm regions. The bar SFE lies in the lower half of the SFE spread in the spiral even though their $\Sigma_{\mathrm{H_{2}}}$ values are comparable, presumably due to the frequent, fast collisions of clouds in the bar region that make the clouds become unbound and thus do not favor star formation (\S\ref{sec_sf} and \ref{sec_disc_gmas}).

\item The SFE variation is also reflected in the relationship between $\Sigma_{\mathrm{H_{2}}}$ and $\Sigma_{\mathrm{SFR}}$, which is the so-called Kennicutt-Schmidt (K-S) relation. The power-law index $N$ decreases from a superlinear relation ($\sim$ 1.7) for the bar GMAs to an almost linear relation ($\sim$ 1) for the spiral and inter-arm GMAs, and to a sublinear relation ($\sim$ 0.8) for the CNR GMAs.  The derived slope cannot be explained by any single star formation mechanism.   A combination of multiple mechanisms or gas phase change is necessary to explain the observed slopes of K-S relation (\S\ref{sec_ks}).
\end{enumerate}

Finally,  we compared the GMA properties obtained from the 12-m observations alone and those obtained from the feathered data (the default data set for this work)(\S\ref{sec_obs_effect}).
Most of the inter-arm GMAs are lost in the 12-m observations due to the missing  extended, low-density structures. 
We identified 70 matched GMAs based on their position and velocity in the two data cubes, and compared the one-to-one relation of GMA properties. 
$M_{\mathrm{c}}$ and $R_{\mathrm{c}}$  decrease in the 12-m-only observation.  On the other hand, removing ACA data has little effect on the median $\sigma_{\mathrm{c}}$, $\Sigma_{\mathrm{H_{2}}}$, and $\alpha_{\mathrm{vir}}$. Moreover,  GMA properties become increasingly uniform in the 12-m observation as suggested by the standard deviation of the GMA properties.  Such uniformity in the properties may  obscure the environmental dependence of the GMA properties.

\section*{Acknowledgement} 
We thank the anonymous referee for providing invaluable comments that helped to improve the quality of the manuscript.
This paper makes use of the following ALMA data: ADS/JAO.ALMA\#2011.0.00004.SV. ALMA is a partnership of ESO (representing its member states), NSF (USA) and NINS (Japan), together with NRC (Canada) and NSC and ASIAA (Taiwan), and KASI (Republic of Korea), in cooperation with the Republic of Chile. The Joint ALMA Observatory is operated by ESO, AUI/NRAO and NAOJ.
This research has made use of the NASA/IPAC Extragalactic Database (NED) which is operated by the Jet Propulsion Laboratory, California Institute of Technology, under contract with the National Aeronautics and Space Administration.
 

\appendix
\begin{longtable*}{llllllllllllllllllll}
	
	\hline
	ID & M$_{\mathrm{c}}$ & R$_{\mathrm{c}}$ & $\sigma_{\mathrm{c}}$ & $\Sigma_{\mathrm{H_{2}}}$ & M$_{\mathrm{vir}}$ & $\alpha_{\mathrm{vir}}$ & $T_{\mathrm{max}}$ & S/N & Type \\
	& (10$^{5}$ M$_{\sun}$) & (pc) & (km s$^{-1}$) & (M$_{\sun}$ pc$^{-2}$) & (10$^{5}$ M$_{\sun}$) & & (K) & &\\  
	(1) & (2) & (3) & (4) & (5) & (6) & (7) & (8) & (9) & (10)\\
	\hline
	1&     50.6$\pm$ 46.7 &      267.0$\pm$206.1 &        3.9$\pm$  3.2 &       22.6$\pm$ 25.9 &        41.8$\pm$ 64.8 &         0.8$\pm$  1.2&         1.0 &      6.7 &    Arm  \\   
	2&    109.4$\pm$ 37.7 &      232.1$\pm$ 58.5 &        7.6$\pm$  1.8 &       64.7$\pm$ 25.7 &       138.3$\pm$ 96.5 &         1.3$\pm$  0.8&         1.3 &      8.7 &    ITA  \\   
	3&     23.2$\pm$ 39.6 &       76.9$\pm$ 47.4 &        4.7$\pm$  8.3 &      125.1$\pm$191.6 &        17.5$\pm$ 76.2 &         0.8$\pm$  2.8&         1.2 &      7.9 &    Arm  \\   
	4&     82.3$\pm$ 37.3 &      438.7$\pm$136.5 &        3.8$\pm$  1.8 &       13.6$\pm$  6.9 &        67.0$\pm$ 73.8 &         0.8$\pm$  0.8&         1.4 &      9.8 &    Arm  \\   
	5&    103.4$\pm$231.0 &      422.7$\pm$500.5 &        5.1$\pm$  9.6 &       18.4$\pm$ 41.1 &       115.9$\pm$472.9 &         1.1$\pm$  4.2&         1.4 &      9.8 &    Arm  \\   
	6&    184.6$\pm$102.8 &      375.5$\pm$354.4 &        8.9$\pm$  5.8 &       41.7$\pm$ 48.2 &       307.1$\pm$640.7 &         1.7$\pm$  2.9&         1.1 &      7.1 &    Bar  \\   
	7&    103.0$\pm$156.2 &      220.5$\pm$161.1 &        4.9$\pm$  5.2 &       67.4$\pm$ 99.0 &        54.5$\pm$121.3 &         0.5$\pm$  1.1&         1.6 &     11.1 &    Arm  \\   
	8&    168.3$\pm$276.1 &      579.6$\pm$259.2 &        7.3$\pm$  5.4 &       15.9$\pm$ 22.4 &       319.5$\pm$546.0 &         1.9$\pm$  3.6&         1.8 &     12.4 &    Arm  \\   
	9&    158.7$\pm$117.6 &      211.8$\pm$108.2 &        9.2$\pm$  4.6 &      112.7$\pm$ 93.3 &       188.1$\pm$169.7 &         1.2$\pm$  1.1&         2.6 &     17.4 &    Arm  \\   
	10&     40.6$\pm$111.3 &      278.2$\pm$213.9 &        7.2$\pm$ 12.0 &       16.7$\pm$ 39.4 &       149.5$\pm$480.3 &         3.7$\pm$ 12.5&         1.2 &      8.3 &    Arm  \\   
	11&    131.9$\pm$ 49.9 &      462.4$\pm$ 77.7 &        6.0$\pm$  2.0 &       19.6$\pm$  7.0 &       171.3$\pm$103.7 &         1.3$\pm$  0.7&         0.9 &      6.1 &    ITA  \\   
	12&     85.9$\pm$ 41.5 &      319.9$\pm$131.6 &        2.9$\pm$  2.9 &       26.7$\pm$ 16.2 &        27.6$\pm$ 48.5 &         0.3$\pm$  0.5&         1.4 &      9.4 &    Arm  \\   
	13&    126.5$\pm$ 96.5 &      285.3$\pm$104.5 &        7.0$\pm$  3.3 &       49.5$\pm$ 36.5 &       144.1$\pm$199.7 &         1.1$\pm$  1.4&         1.2 &      8.5 &    Arm  \\   
	14&     42.7$\pm$ 66.5 &      178.8$\pm$138.2 &       10.1$\pm$  7.2 &       42.5$\pm$ 64.7 &       191.5$\pm$178.2 &         4.5$\pm$  6.5&         1.5 &     10.4 &    Arm  \\   
	15&     96.9$\pm$ 63.6 &      345.0$\pm$114.3 &        3.1$\pm$  3.5 &       25.9$\pm$ 16.7 &        33.4$\pm$ 73.0 &         0.3$\pm$  0.6&         1.6 &     10.6 &    Arm  \\   
	16&    149.3$\pm$151.3 &      313.5$\pm$241.0 &        6.6$\pm$  6.1 &       48.3$\pm$ 57.5 &       142.7$\pm$220.8 &         1.0$\pm$  1.4&         1.4 &      9.5 &    Bar  \\   
	17&    126.3$\pm$ 22.8 &      468.4$\pm$ 76.1 &        5.4$\pm$  2.2 &       18.3$\pm$  4.3 &       140.5$\pm$105.0 &         1.1$\pm$  0.7&         1.4 &      9.4 &    ITA  \\   
	18&     42.2$\pm$ 46.6 &      166.3$\pm$137.0 &        4.9$\pm$  4.3 &       48.6$\pm$ 62.4 &        40.8$\pm$ 73.0 &         1.0$\pm$  1.6&         1.4 &      9.2 &    Arm  \\   
	19&    211.3$\pm$218.4 &      524.2$\pm$457.5 &       14.3$\pm$ 17.2 &       24.5$\pm$ 31.5 &       1110.0$\pm$2618.7 &         5.3$\pm$ 10.8&         1.8 &     12.1 &    Arm  \\   
	20&     58.7$\pm$ 98.0 &      230.2$\pm$176.6 &        8.8$\pm$ 10.2 &       35.3$\pm$ 56.2 &       187.5$\pm$418.2 &         3.2$\pm$  7.1&         1.8 &     12.2 &    Arm  \\   
	21&     43.0$\pm$ 38.3 &      202.4$\pm$122.6 &        8.5$\pm$  6.3 &       33.4$\pm$ 33.0 &       153.2$\pm$255.0 &         3.6$\pm$  5.4&         1.5 &     10.1 &    Arm  \\   
	22&    398.6$\pm$417.4 &      380.4$\pm$204.8 &       12.9$\pm$  8.5 &       87.7$\pm$ 90.8 &       660.5$\pm$1031.4 &         1.7$\pm$  2.5&         3.8 &     26.1 &   CNR   \\   
	23&     59.7$\pm$ 23.5 &      312.3$\pm$ 77.7 &        4.9$\pm$  3.4 &       19.5$\pm$  8.2 &        78.0$\pm$112.3 &         1.3$\pm$  1.6&         1.7 &     11.7 &    Arm  \\   
	24&     39.3$\pm$ 20.2 &      154.0$\pm$ 83.4 &        4.1$\pm$  3.5 &       52.7$\pm$ 38.9 &        27.3$\pm$ 47.2 &         0.7$\pm$  1.0&         1.8 &     12.4 &    Arm  \\   
	25&     26.1$\pm$ 57.6 &      146.3$\pm$238.7 &        1.6$\pm$  1.4 &       38.8$\pm$ 99.1 &         3.8$\pm$  9.6 &         0.1$\pm$  0.4&         0.9 &      5.9 &    Arm  \\   
	26&    218.7$\pm$119.1 &      298.2$\pm$124.1 &        6.6$\pm$  2.9 &       78.3$\pm$ 50.2 &       136.1$\pm$147.6 &         0.6$\pm$  0.6&         2.1 &     13.9 &    Arm  \\   
	27&     73.9$\pm$ 28.4 &      179.8$\pm$101.7 &        7.6$\pm$  3.0 &       72.7$\pm$ 51.6 &       108.6$\pm$ 90.8 &         1.5$\pm$  1.1&         1.5 &     10.2 &    Arm  \\   
	28&     29.9$\pm$ 62.9 &      263.5$\pm$194.6 &        3.4$\pm$  6.8 &       13.7$\pm$ 25.8 &        31.3$\pm$166.1 &         1.0$\pm$  4.8&         1.0 &      6.5 &    Arm  \\   
	29&     55.5$\pm$ 47.5 &      212.1$\pm$113.7 &        6.5$\pm$  3.1 &       39.3$\pm$ 35.9 &        92.7$\pm$111.5 &         1.7$\pm$  2.0&         1.7 &     11.6 &    ITA  \\   
	30&     27.1$\pm$ 24.4 &      244.4$\pm$345.3 &        7.7$\pm$  2.3 &       14.5$\pm$ 25.3 &       150.0$\pm$232.1 &         5.5$\pm$  7.9&         0.9 &      6.0 &    Arm  \\   
	31&    107.2$\pm$ 52.3 &      553.0$\pm$127.9 &       15.4$\pm$  5.3 &       11.2$\pm$  5.2 &       1366.0$\pm$1206.5 &        12.7$\pm$ 10.3&         0.9 &      5.9 &    Arm  \\   
	32&    467.2$\pm$409.6 &      540.7$\pm$680.0 &       19.8$\pm$ 16.1 &       50.9$\pm$ 80.7 &       2209.0$\pm$6950.4 &         4.7$\pm$ 12.4&         3.4 &     22.9 &    CNR   \\   
	33&     45.8$\pm$ 98.7 &      232.4$\pm$195.6 &        6.7$\pm$  8.1 &       27.0$\pm$ 53.2 &       109.8$\pm$353.6 &         2.4$\pm$  7.4&         1.4 &      9.4 &    ITA  \\   
	34&     53.4$\pm$ 73.1 &      240.6$\pm$213.6 &       11.1$\pm$  7.3 &       29.4$\pm$ 43.7 &       307.8$\pm$344.2 &         5.8$\pm$  8.1&         1.6 &     11.1 &    Arm  \\   
	35&    306.4$\pm$306.6 &      341.4$\pm$227.0 &        6.5$\pm$  3.8 &       83.7$\pm$ 91.9 &       150.3$\pm$248.7 &         0.5$\pm$  0.8&         1.7 &     11.6 &    Arm  \\   
	36&    109.1$\pm$112.6 &      322.8$\pm$303.7 &        6.8$\pm$  6.3 &       33.3$\pm$ 44.9 &       155.4$\pm$413.7 &         1.4$\pm$  3.3&         1.3 &      8.8 &    Arm  \\   
	37&     72.2$\pm$ 71.6 &      195.7$\pm$152.6 &       12.1$\pm$  7.6 &       60.0$\pm$ 71.2 &       295.6$\pm$647.7 &         4.1$\pm$  7.9&         2.5 &     16.7 &    Bar  \\   
	38&     40.9$\pm$ 36.2 &      199.7$\pm$ 87.5 &        3.9$\pm$  1.6 &       32.7$\pm$ 28.2 &        32.2$\pm$ 29.6 &         0.8$\pm$  0.8&         0.8 &      5.2 &    ITA  \\   
	39&     21.1$\pm$ 17.4 &      213.4$\pm$103.8 &        5.9$\pm$  3.8 &       14.8$\pm$ 12.7 &        78.1$\pm$102.7 &         3.7$\pm$  4.6&         0.8 &      5.2 &    ITA  \\   
	40&    433.5$\pm$279.9 &      353.4$\pm$130.3 &       23.6$\pm$ 11.1 &      110.5$\pm$ 73.4 &       2054.0$\pm$2040.9 &         4.7$\pm$  4.5&         3.5 &     23.9 &    Bar  \\   
	41&     24.2$\pm$ 22.8 &      222.4$\pm$171.8 &       17.2$\pm$ 10.2 &       15.6$\pm$ 18.0 &       680.9$\pm$1047.5 &        28.1$\pm$ 40.5&         1.0 &      7.0 &    Arm  \\   
	42&     68.2$\pm$ 73.7 &      236.0$\pm$196.3 &        3.1$\pm$  2.9 &       39.0$\pm$ 49.8 &        23.8$\pm$ 56.3 &         0.3$\pm$  0.7&         1.2 &      7.9 &    Arm  \\   
	43&    139.8$\pm$325.5 &      301.7$\pm$215.5 &       13.0$\pm$ 16.1 &       48.9$\pm$ 99.3 &       526.3$\pm$1272.8 &         3.8$\pm$ 10.1&         3.1 &     21.3 &    CNR   \\   
	44&    844.6$\pm$377.7 &      471.1$\pm$140.6 &       15.8$\pm$  3.4 &      121.2$\pm$ 59.6 &       1217.0$\pm$669.8 &         1.4$\pm$  0.8&         3.2 &     21.5 &    Bar  \\   
	45&    255.5$\pm$349.9 &      466.5$\pm$226.2 &       10.2$\pm$  5.0 &       37.4$\pm$ 45.8 &       500.8$\pm$580.9 &         2.0$\pm$  2.8&         2.0 &     13.5 &    Bar  \\   
	46&     82.2$\pm$112.3 &      392.7$\pm$340.0 &        6.3$\pm$  6.6 &       17.0$\pm$ 24.9 &       163.1$\pm$318.9 &         2.0$\pm$  3.8&         1.2 &      7.9 &    ITA  \\   
	47&     28.1$\pm$ 23.1 &      185.3$\pm$ 99.9 &        3.9$\pm$  3.8 &       26.1$\pm$ 23.4 &        29.1$\pm$ 48.8 &         1.0$\pm$  1.5&         1.0 &      7.0 &    Arm  \\   
	48&    336.1$\pm$554.2 &      229.4$\pm$258.9 &        7.3$\pm$ 29.3 &      203.3$\pm$373.3 &       126.3$\pm$885.2 &         0.4$\pm$  2.2&         4.3 &     28.9 &    CNR   \\   
	49&     24.5$\pm$ 22.0 &      158.9$\pm$133.4 &        7.1$\pm$  3.5 &       30.9$\pm$ 36.8 &        82.4$\pm$102.1 &         3.4$\pm$  4.1&         0.8 &      5.5 &    ITA  \\   
	50&     39.2$\pm$ 42.0 &      288.8$\pm$193.1 &        5.4$\pm$  4.6 &       15.0$\pm$ 17.1 &        88.4$\pm$132.6 &         2.3$\pm$  3.3&         0.8 &      5.7 &    Arm  \\   
	51&     45.4$\pm$ 32.9 &      248.4$\pm$ 67.6 &       12.9$\pm$  6.9 &       23.4$\pm$ 15.4 &       426.7$\pm$497.7 &         9.4$\pm$ 10.3&         1.1 &      7.3 &    ITA  \\   
	52&    208.6$\pm$ 45.6 &      426.0$\pm$ 58.6 &        8.2$\pm$  1.4 &       36.6$\pm$  8.6 &       299.2$\pm$115.1 &         1.4$\pm$  0.5&         1.6 &     11.0 &    Arm  \\   
	53&    164.8$\pm$ 89.4 &      286.6$\pm$141.0 &       12.2$\pm$  4.5 &       63.9$\pm$ 45.1 &       445.1$\pm$429.4 &         2.7$\pm$  2.4&         2.5 &     16.7 &    Arm  \\   
	54&    150.7$\pm$201.7 &      285.5$\pm$342.6 &        5.9$\pm$  5.8 &       58.9$\pm$101.8 &       104.4$\pm$308.7 &         0.7$\pm$  1.8&         1.7 &     11.3 &    Arm  \\   
	55&     19.0$\pm$ 24.9 &      189.3$\pm$156.1 &        9.7$\pm$  6.5 &       16.9$\pm$ 23.7 &       184.8$\pm$325.2 &         9.7$\pm$ 17.1&         0.8 &      5.7 &    Bar  \\   
	56&     63.5$\pm$ 35.9 &      425.7$\pm$158.3 &       10.8$\pm$  4.3 &       11.2$\pm$  6.9 &       514.5$\pm$404.2 &         8.1$\pm$  6.3&         0.9 &      6.3 &    ITA  \\   
	57&    128.6$\pm$ 87.3 &      325.3$\pm$277.2 &        5.0$\pm$  4.9 &       38.7$\pm$ 42.8 &        85.7$\pm$201.6 &         0.7$\pm$  1.3&         1.1 &      7.4 &    ITA  \\   
	58&     72.2$\pm$173.7 &      225.3$\pm$161.7 &        3.0$\pm$  6.2 &       45.3$\pm$ 94.6 &        20.9$\pm$ 88.1 &         0.3$\pm$  1.1&         1.9 &     13.0 &    Arm  \\   
	59&    150.1$\pm$201.1 &      174.8$\pm$181.8 &       21.7$\pm$ 13.7 &      156.3$\pm$248.8 &       859.2$\pm$1253.7 &         5.7$\pm$  9.1&         3.3 &     22.4 &    CNR   \\   
	60&    278.2$\pm$132.6 &      600.3$\pm$198.8 &       13.9$\pm$ 10.6 &       24.6$\pm$ 13.1 &       1215.0$\pm$2061.6 &         4.4$\pm$  6.2&         2.2 &     15.0 &    Arm  \\   
	61&    111.0$\pm$192.6 &      253.6$\pm$156.2 &        4.2$\pm$  3.9 &       54.9$\pm$ 85.3 &        46.0$\pm$ 78.1 &         0.4$\pm$  0.8&         0.9 &      6.4 &    Bar  \\   
	62&    231.9$\pm$ 69.2 &      505.3$\pm$112.4 &        7.5$\pm$  2.1 &       28.9$\pm$ 10.0 &       294.9$\pm$161.1 &         1.3$\pm$  0.6&         1.3 &      8.5 &    ITA  \\   
	63&     37.8$\pm$ 79.1 &      173.7$\pm$186.8 &        1.4$\pm$  5.0 &       39.9$\pm$ 82.5 &         3.6$\pm$ 20.5 &         0.1$\pm$  0.5&         0.9 &      5.8 &    ITA  \\   
	64&    814.5$\pm$662.7 &      138.9$\pm$ 55.9 &       52.3$\pm$ 18.2 &      1343.9$\pm$1067.5 &       3949.0$\pm$3939.5 &         4.8$\pm$  5.0&         4.2 &     28.8 &    CNR   \\   
	65&     34.0$\pm$ 27.4 &      258.8$\pm$156.9 &        9.3$\pm$  3.2 &       16.1$\pm$ 15.2 &       233.3$\pm$313.2 &         6.9$\pm$  8.6&         0.7 &      5.1 &    Bar  \\   
	66&     69.0$\pm$ 72.9 &      462.3$\pm$254.4 &        3.3$\pm$  4.4 &       10.3$\pm$ 10.8 &        52.7$\pm$129.5 &         0.8$\pm$  1.6&         0.9 &      5.9 &    Arm  \\   
	67&    567.0$\pm$557.5 &      318.3$\pm$174.4 &       26.5$\pm$ 14.9 &      178.1$\pm$178.4 &       2921.4$\pm$4695.2 &         5.2$\pm$  7.8&         4.5 &     30.6 &    CNR   \\   
	68&     84.6$\pm$ 70.7 &      552.2$\pm$242.1 &        2.0$\pm$  2.8 &        8.8$\pm$  7.4 &        24.0$\pm$ 64.0 &         0.3$\pm$  0.6&         1.2 &      8.0 &    ITA  \\   
	69&    109.7$\pm$ 25.5 &      401.7$\pm$ 95.8 &        6.4$\pm$  1.7 &       21.6$\pm$  7.1 &       171.2$\pm$139.4 &         1.6$\pm$  1.1&         1.7 &     11.5 &    ITA  \\   
	70&    191.3$\pm$171.9 &      263.8$\pm$123.2 &        9.8$\pm$  5.5 &       87.5$\pm$ 78.1 &       265.3$\pm$492.6 &         1.4$\pm$  2.3&         2.3 &     15.5 &    Arm  \\   
	71&    103.0$\pm$ 67.2 &      385.7$\pm$150.3 &        8.1$\pm$  5.3 &       22.0$\pm$ 15.0 &       265.7$\pm$416.0 &         2.6$\pm$  3.5&         1.0 &      6.5 &    Bar  \\   
	72&     81.9$\pm$ 34.4 &      174.9$\pm$ 77.1 &        8.2$\pm$  2.7 &       85.1$\pm$ 51.2 &       121.2$\pm$ 99.6 &         1.5$\pm$  1.1&         1.4 &      9.2 &    Bar  \\   
	73&    261.1$\pm$ 84.6 &      430.0$\pm$142.8 &        7.2$\pm$  2.0 &       45.0$\pm$ 20.5 &       232.1$\pm$166.2 &         0.9$\pm$  0.6&         1.7 &     11.3 &    Arm  \\   
	74&     78.2$\pm$149.9 &      280.5$\pm$162.3 &        8.0$\pm$  8.6 &       31.6$\pm$ 52.7 &       187.8$\pm$376.7 &         2.4$\pm$  5.3&         2.0 &     13.4 &    Arm  \\   
	75&    188.2$\pm$149.4 &      348.0$\pm$138.1 &       14.4$\pm$  4.0 &       49.5$\pm$ 38.5 &       754.7$\pm$580.2 &         4.0$\pm$  3.5&         2.0 &     13.4 &    Arm  \\   
	76&     28.2$\pm$ 24.8 &      243.7$\pm$212.7 &        9.1$\pm$  6.5 &       15.1$\pm$ 18.3 &       209.0$\pm$311.8 &         7.4$\pm$ 10.3&         0.9 &      5.8 &    Bar  \\   
	77&    109.1$\pm$ 45.4 &      337.5$\pm$181.7 &       15.6$\pm$  6.4 &       30.5$\pm$ 21.2 &       856.9$\pm$854.2 &         7.9$\pm$  6.8&         1.0 &      6.9 &    Bar  \\   
	78&     66.3$\pm$ 47.5 &      380.0$\pm$258.1 &        8.8$\pm$  5.7 &       14.6$\pm$ 14.0 &       308.6$\pm$534.2 &         4.7$\pm$  7.0&         1.7 &     11.7 &    Arm  \\   
	79&    161.6$\pm$ 75.2 &      442.1$\pm$139.3 &        8.0$\pm$  4.5 &       26.3$\pm$ 13.6 &       297.1$\pm$402.6 &         1.8$\pm$  2.1&         1.5 &      9.9 &    ITA  \\   
	80&     28.5$\pm$ 75.7 &      238.6$\pm$151.6 &        5.9$\pm$  9.9 &       16.0$\pm$ 35.7 &        86.0$\pm$305.9 &         3.0$\pm$ 10.7&         1.1 &      7.7 &    ITA  \\   
	81&     48.2$\pm$ 97.5 &      201.0$\pm$206.6 &        9.8$\pm$  8.5 &       38.0$\pm$ 75.7 &       200.7$\pm$452.5 &         4.2$\pm$ 10.1&         1.2 &      8.2 &    ITA  \\   
	82&    102.0$\pm$ 73.4 &      317.1$\pm$358.7 &       14.4$\pm$ 23.2 &       32.3$\pm$ 45.3 &       685.7$\pm$2703.3 &         6.7$\pm$ 21.6&         1.1 &      7.6 &    Bar  \\   
	83&    159.7$\pm$127.0 &      414.5$\pm$320.4 &        5.1$\pm$  3.6 &       29.6$\pm$ 32.0 &       111.0$\pm$224.2 &         0.7$\pm$  1.2&         1.7 &     11.7 &    Arm  \\   
	84&    144.6$\pm$116.1 &      324.4$\pm$ 97.1 &        6.3$\pm$  3.7 &       43.7$\pm$ 31.8 &       133.0$\pm$195.9 &         0.9$\pm$  1.2&         2.3 &     15.5 &    Arm  \\   
	85&    1024.0$\pm$867.5 &      328.2$\pm$253.9 &       31.0$\pm$ 20.8 &      302.6$\pm$335.0 &       3278.0$\pm$3705.5 &         3.2$\pm$  3.6&         4.5 &     30.5 &    CNR   \\   
	86&    373.5$\pm$261.4 &      602.2$\pm$218.7 &        7.5$\pm$  4.9 &       32.8$\pm$ 22.8 &       354.3$\pm$549.3 &         0.9$\pm$  1.3&         1.9 &     12.7 &    Arm  \\   
	87&    281.7$\pm$338.7 &      232.1$\pm$256.5 &       30.8$\pm$ 30.3 &      166.4$\pm$262.5 &       2289.0$\pm$4909.4 &         8.1$\pm$ 16.0&         3.9 &     26.7 &    CNR   \\   
	88&     69.5$\pm$ 86.0 &      282.1$\pm$129.1 &        8.5$\pm$  5.1 &       27.8$\pm$ 31.1 &       213.8$\pm$386.6 &         3.1$\pm$  5.4&         2.2 &     14.6 &    Arm  \\   
	89&    234.8$\pm$380.0 &      232.6$\pm$319.3 &        1.3$\pm$  4.9 &      138.1$\pm$279.3 &         4.3$\pm$ 25.4 &         0.0$\pm$  0.1&         1.6 &     11.1 &    Bar  \\   
	90&    166.1$\pm$ 80.4 &      349.3$\pm$128.5 &       10.6$\pm$  5.9 &       43.3$\pm$ 24.6 &       409.3$\pm$491.2 &         2.5$\pm$  2.6&         2.4 &     16.0 &    Arm  \\   
	91&    114.0$\pm$123.6 &      207.2$\pm$212.2 &       13.8$\pm$ 15.6 &       84.5$\pm$122.3 &       410.5$\pm$931.7 &         3.6$\pm$  7.2&         1.9 &     13.1 &    Bar  \\   
	92&    100.9$\pm$165.0 &      498.2$\pm$399.0 &        9.8$\pm$ 15.1 &       12.9$\pm$ 20.6 &       493.2$\pm$1725.0 &         4.9$\pm$ 15.1&         1.9 &     12.9 &    Bar  \\   
	93&    124.4$\pm$ 68.2 &      403.9$\pm$106.3 &        6.1$\pm$  4.2 &       24.3$\pm$ 12.9 &       154.0$\pm$238.4 &         1.2$\pm$  1.6&         1.5 &     10.3 &    Bar  \\   
	94&    322.9$\pm$322.6 &      524.5$\pm$193.9 &       17.5$\pm$  6.2 &       37.4$\pm$ 33.7 &       1665.0$\pm$1875.5 &         5.2$\pm$  6.2&         2.2 &     15.0 &    Arm  \\   
	95&    330.9$\pm$168.1 &      431.9$\pm$110.9 &        7.4$\pm$  2.0 &       56.5$\pm$ 28.2 &       246.8$\pm$149.9 &         0.7$\pm$  0.5&         2.4 &     16.6 &    Arm  \\   
	96&    124.6$\pm$ 92.6 &      502.8$\pm$372.0 &        8.6$\pm$  3.7 &       15.7$\pm$ 16.1 &       382.8$\pm$392.6 &         3.1$\pm$  3.1&         1.1 &      7.7 &    ITA  \\   
	97&    249.9$\pm$139.7 &      381.5$\pm$119.9 &       12.1$\pm$  4.6 &       54.7$\pm$ 31.2 &       579.9$\pm$574.8 &         2.3$\pm$  2.1&         2.9 &     19.9 &    Arm  \\   
	98&     97.7$\pm$ 54.5 &      244.8$\pm$107.5 &        9.8$\pm$  3.2 &       51.9$\pm$ 34.6 &       244.3$\pm$195.6 &         2.5$\pm$  2.0&         1.6 &     10.7 &    Bar  \\   
	99&    182.8$\pm$221.7 &      396.9$\pm$191.5 &        4.2$\pm$  3.0 &       36.9$\pm$ 41.1 &        72.5$\pm$149.4 &         0.4$\pm$  0.8&         1.7 &     11.7 &    Arm  \\   
	100&     59.3$\pm$ 60.9 &      232.3$\pm$ 98.9 &        4.6$\pm$  2.6 &       35.0$\pm$ 33.3 &        51.2$\pm$ 41.2 &         0.9$\pm$  0.9&         1.1 &      7.6 &    ITA  \\   
	101&    161.1$\pm$ 91.6 &      312.5$\pm$126.2 &        5.8$\pm$  3.1 &       52.5$\pm$ 33.9 &       110.7$\pm$139.0 &         0.7$\pm$  0.8&         2.3 &     15.4 &    Arm  \\   
	102&    114.8$\pm$ 39.5 &      566.6$\pm$195.8 &        7.5$\pm$  2.0 &       11.4$\pm$  5.4 &       329.1$\pm$283.3 &         2.9$\pm$  2.1&         1.5 &     10.3 &    ITA  \\   
	103&    106.4$\pm$113.8 &      218.3$\pm$118.6 &       13.0$\pm$  7.6 &       71.1$\pm$ 74.9 &       384.8$\pm$477.2 &         3.6$\pm$  4.7&         1.7 &     11.4 &    Arm  \\   
	104&     37.3$\pm$ 41.4 &      130.3$\pm$111.1 &       11.5$\pm$  6.0 &       69.9$\pm$ 91.6 &       179.9$\pm$240.6 &         4.8$\pm$  6.7&         1.4 &      9.3 &    Bar  \\   
	105&     77.5$\pm$113.1 &      269.3$\pm$173.8 &        8.4$\pm$  8.0 &       34.0$\pm$ 46.9 &       196.7$\pm$499.1 &         2.5$\pm$  5.9&         1.9 &     12.9 &    Bar  \\   
	106&     92.8$\pm$101.9 &      238.5$\pm$103.8 &        4.7$\pm$  4.6 &       51.9$\pm$ 52.3 &        54.6$\pm$106.0 &         0.6$\pm$  1.1&         1.9 &     12.9 &    Arm  \\   
	107&    323.8$\pm$ 81.1 &      359.5$\pm$ 87.7 &        7.7$\pm$  2.5 &       79.8$\pm$ 27.2 &       222.0$\pm$106.0 &         0.7$\pm$  0.3&         3.4 &     22.9 &    Arm  \\   
	108&    126.7$\pm$ 71.2 &      237.8$\pm$117.7 &       14.8$\pm$  4.3 &       71.3$\pm$ 51.2 &       538.4$\pm$537.5 &         4.2$\pm$  3.9&         1.3 &      8.8 &    ITA  \\   
	109&    751.4$\pm$ 64.9 &      422.2$\pm$ 63.8 &       10.0$\pm$  1.5 &      134.2$\pm$ 24.8 &       435.9$\pm$137.4 &         0.6$\pm$  0.2&         3.7 &     25.4 &    Arm  \\   
	110&     81.3$\pm$ 53.9 &      258.4$\pm$102.7 &        4.9$\pm$  2.3 &       38.8$\pm$ 27.0 &        64.5$\pm$ 54.3 &         0.8$\pm$  0.7&         1.1 &      7.6 &    Arm  \\   
	111&     76.2$\pm$ 39.7 &      506.8$\pm$234.3 &       10.8$\pm$  4.4 &        9.4$\pm$  6.3 &       619.6$\pm$718.2 &         8.1$\pm$  8.3&         1.6 &     10.8 &    ITA  \\   
	112&     28.1$\pm$ 16.3 &      191.3$\pm$130.8 &        6.3$\pm$  2.2 &       24.5$\pm$ 22.1 &        78.9$\pm$ 80.3 &         2.8$\pm$  2.6&         0.7 &      5.0 &    ITA  \\   
	113&    152.4$\pm$117.3 &      437.3$\pm$173.5 &       10.2$\pm$  3.4 &       25.4$\pm$ 19.3 &       472.7$\pm$375.5 &         3.1$\pm$  2.7&         2.4 &     16.3 &    Arm  \\   
	114&    155.0$\pm$113.8 &      211.6$\pm$ 73.5 &        2.5$\pm$  3.4 &      110.2$\pm$ 77.9 &        14.1$\pm$ 38.7 &         0.1$\pm$  0.2&         1.5 &     10.2 &    ITA  \\   
	115&    1543.0$\pm$1930.6 &      321.3$\pm$296.6 &        8.7$\pm$ 15.9 &      475.7$\pm$688.3 &       253.5$\pm$1122.9 &         0.2$\pm$  0.6&         4.5 &     30.3 &    CNR   \\   
	116&    872.6$\pm$239.4 &      253.3$\pm$ 56.1 &       12.3$\pm$  3.1 &      433.0$\pm$144.3 &       396.9$\pm$248.9 &         0.5$\pm$  0.2&         6.9 &     46.7 &    CNR   \\   
	117&     57.4$\pm$ 44.3 &      412.5$\pm$142.9 &        1.7$\pm$  0.7 &       10.7$\pm$  7.9 &        12.5$\pm$ 11.9 &         0.2$\pm$  0.2&         0.9 &      5.9 &    ITA  \\   
	118&     23.0$\pm$ 22.1 &      342.5$\pm$160.3 &        2.5$\pm$  5.0 &        6.2$\pm$  5.8 &        21.7$\pm$ 90.9 &         0.9$\pm$  3.2&         0.9 &      6.1 &    ITA  \\   
	119&     80.9$\pm$ 50.2 &      594.6$\pm$163.6 &       10.8$\pm$  4.4 &        7.3$\pm$  4.3 &       721.1$\pm$760.9 &         8.9$\pm$  8.7&         1.1 &      7.2 &    ITA  \\   
	120&     88.1$\pm$130.2 &      499.7$\pm$255.4 &       10.0$\pm$  7.4 &       11.2$\pm$ 14.8 &       522.6$\pm$864.6 &         5.9$\pm$ 10.5&         1.6 &     11.0 &    ITA  \\   
	121&     52.8$\pm$ 36.9 &      274.6$\pm$260.1 &        4.4$\pm$  4.1 &       22.3$\pm$ 27.0 &        55.1$\pm$134.9 &         1.0$\pm$  2.1&         0.8 &      5.3 &    ITA  \\   
	122&     81.5$\pm$ 13.7 &      520.7$\pm$ 67.1 &        7.2$\pm$  2.4 &        9.6$\pm$  1.9 &       281.7$\pm$196.1 &         3.5$\pm$  2.0&         0.9 &      6.2 &    ITA  \\   
	123&     79.8$\pm$114.7 &      145.5$\pm$135.3 &        6.4$\pm$  3.3 &      119.9$\pm$187.0 &        61.8$\pm$105.2 &         0.8$\pm$  1.4&         1.2 &      8.2 &    Arm  \\   
	124&    125.7$\pm$168.1 &      155.8$\pm$165.2 &       10.5$\pm$  5.5 &      164.7$\pm$264.8 &       178.8$\pm$283.8 &         1.4$\pm$  2.4&         1.8 &     12.2 &    ITA  \\   
	125&    425.7$\pm$409.0 &      315.0$\pm$122.2 &        8.9$\pm$  4.7 &      136.5$\pm$120.9 &       258.8$\pm$310.6 &         0.6$\pm$  0.7&         3.4 &     22.9 &    Arm  \\   
	126&     73.2$\pm$134.5 &      315.9$\pm$146.6 &        3.3$\pm$  5.8 &       23.3$\pm$ 36.4 &        35.7$\pm$130.3 &         0.5$\pm$  1.6&         1.1 &      7.5 &    ITA  \\   
	127&     53.0$\pm$ 42.2 &      466.7$\pm$282.3 &        8.2$\pm$  4.2 &        7.7$\pm$  7.2 &       330.3$\pm$501.8 &         6.2$\pm$  8.6&         0.8 &      5.7 &    ITA  \\   
	128&     86.7$\pm$ 61.8 &      153.7$\pm$ 68.8 &        5.8$\pm$  2.6 &      116.8$\pm$ 89.2 &        54.3$\pm$ 49.0 &         0.6$\pm$  0.6&         1.6 &     10.8 &    ITA  \\   
	129&    283.2$\pm$ 80.4 &      748.1$\pm$189.7 &        5.4$\pm$  1.5 &       16.1$\pm$  5.9 &       223.7$\pm$173.2 &         0.8$\pm$  0.5&         1.4 &      9.7 &    ITA  \\   
	130&     35.4$\pm$ 51.4 &      396.9$\pm$121.6 &        5.4$\pm$  3.7 &        7.1$\pm$  8.7 &       120.5$\pm$195.0 &         3.4$\pm$  5.9&         0.8 &      5.7 &    ITA  \\   
	131&     58.2$\pm$ 71.9 &      314.4$\pm$147.7 &        3.3$\pm$  2.2 &       18.7$\pm$ 21.0 &        36.3$\pm$ 58.1 &         0.6$\pm$  1.0&         0.8 &      5.2 &    ITA  \\   
	132&     33.1$\pm$ 14.3 &      152.0$\pm$ 93.7 &        9.9$\pm$  6.1 &       45.5$\pm$ 35.4 &       154.4$\pm$196.9 &         4.7$\pm$  5.0&         1.0 &      6.8 &    ITA  \\   
	133&     18.2$\pm$ 18.5 &      205.7$\pm$164.0 &       14.9$\pm$ 14.9 &       13.7$\pm$ 16.7 &       475.7$\pm$1115.4 &        26.1$\pm$ 53.3&         0.8 &      5.3 &    ITA  \\   
	134&    224.1$\pm$169.1 &      372.4$\pm$603.9 &        4.1$\pm$  7.3 &       51.4$\pm$ 99.3 &        65.2$\pm$307.6 &         0.3$\pm$  1.1&         2.1 &     14.4 &    ITA  \\   
	135&    124.2$\pm$ 29.1 &      497.9$\pm$ 82.5 &        6.5$\pm$  1.2 &       15.9$\pm$  4.2 &       221.8$\pm$110.0 &         1.8$\pm$  0.8&         0.9 &      6.2 &    ITA  \\   
	136&     30.7$\pm$ 20.5 &      332.2$\pm$184.4 &        5.2$\pm$  2.6 &        8.8$\pm$  7.3 &        92.9$\pm$ 97.1 &         3.0$\pm$  3.0&         0.8 &      5.4 &    ITA  \\   
	137&    102.0$\pm$ 62.4 &      298.3$\pm$113.4 &       11.0$\pm$  5.1 &       36.5$\pm$ 23.8 &       372.6$\pm$456.4 &         3.7$\pm$  4.0&         1.5 &     10.2 &    ITA  \\   
	138&     33.1$\pm$ 16.9 &      227.6$\pm$118.9 &        9.2$\pm$  3.5 &       20.4$\pm$ 14.6 &       201.3$\pm$244.8 &         6.1$\pm$  6.4&         0.8 &      5.7 &    ITA  \\   
	139&     61.0$\pm$ 49.3 &      400.6$\pm$156.1 &        4.7$\pm$  2.7 &       12.1$\pm$  9.5 &        91.5$\pm$136.2 &         1.5$\pm$  2.0&         1.2 &      8.0 &    Arm  \\   
	140&     55.3$\pm$ 31.3 &      367.4$\pm$144.9 &        5.0$\pm$  2.3 &       13.0$\pm$  8.3 &        94.6$\pm$124.7 &         1.7$\pm$  2.0&         0.8 &      5.4 &    ITA  \\   
	141&     73.1$\pm$ 11.8 &      276.6$\pm$ 60.9 &        8.7$\pm$  1.9 &       30.4$\pm$  8.5 &       220.1$\pm$132.9 &         3.0$\pm$  1.5&         0.8 &      5.6 &    ITA  \\   
	142&     41.1$\pm$ 24.1 &      164.5$\pm$ 98.5 &        3.7$\pm$  3.7 &       48.4$\pm$ 39.8 &        23.7$\pm$ 56.6 &         0.6$\pm$  1.1&         1.0 &      6.7 &    ITA  \\   
	143&     49.8$\pm$ 23.4 &      282.9$\pm$298.3 &        6.1$\pm$  3.0 &       19.8$\pm$ 24.8 &       108.3$\pm$189.2 &         2.2$\pm$  3.1&         0.9 &      6.4 &    ITA  \\   
	144&    163.1$\pm$ 36.5 &      552.1$\pm$131.2 &        9.7$\pm$  2.9 &       17.0$\pm$  5.5 &       544.9$\pm$479.5 &         3.3$\pm$  2.4&         1.0 &      7.0 &    ITA  \\   
	145&     81.1$\pm$ 41.7 &      242.3$\pm$ 59.3 &        5.7$\pm$  1.5 &       44.0$\pm$ 21.8 &        83.1$\pm$ 50.0 &         1.0$\pm$  0.6&         0.9 &      6.0 &    ITA  \\   
	146&     48.5$\pm$ 27.9 &      279.6$\pm$ 94.6 &        6.4$\pm$  2.7 &       19.8$\pm$ 11.8 &       118.4$\pm$126.2 &         2.4$\pm$  2.4&         0.8 &      5.7 &    ITA  \\   
	147&    106.9$\pm$  8.8 &      553.7$\pm$ 63.3 &        6.1$\pm$  0.9 &       11.1$\pm$  1.6 &       217.0$\pm$ 77.6 &         2.0$\pm$  0.6&         1.1 &      7.6 &    ITA  \\   
	148&    177.2$\pm$ 21.3 &      678.4$\pm$ 61.3 &        9.9$\pm$  2.7 &       12.3$\pm$  1.7 &       698.5$\pm$390.0 &         3.9$\pm$  1.8&         0.8 &      5.6 &    ITA  \\   
	149&     22.0$\pm$ 15.7 &      193.8$\pm$113.2 &        6.2$\pm$  5.8 &       18.7$\pm$ 16.3 &        78.0$\pm$141.7 &         3.5$\pm$  5.5&         0.9 &      6.2 &    ITA  \\   
	150&     23.2$\pm$ 22.8 &      167.8$\pm$ 92.5 &        6.5$\pm$  3.1 &       26.2$\pm$ 26.3 &        74.4$\pm$ 98.2 &         3.2$\pm$  4.2&         0.8 &      5.2 &    ITA  \\   
	151&     50.2$\pm$ 50.3 &      269.1$\pm$155.5 &        1.8$\pm$  2.7 &       22.1$\pm$ 22.8 &         9.2$\pm$ 24.4 &         0.2$\pm$  0.4&         0.9 &      6.1 &    ITA  \\   
	152&    110.7$\pm$  8.3 &      420.5$\pm$ 38.0 &        6.3$\pm$  1.0 &       19.9$\pm$  2.4 &       172.6$\pm$ 69.2 &         1.6$\pm$  0.5&         1.0 &      6.7 &    ITA  \\   
	153&     94.2$\pm$ 30.1 &      397.4$\pm$ 92.5 &        7.4$\pm$  3.0 &       19.0$\pm$  7.0 &       225.0$\pm$157.3 &         2.4$\pm$  1.5&         1.1 &      7.1 &    ITA  \\   
	154&     23.1$\pm$ 13.5 &      168.9$\pm$166.4 &        6.9$\pm$  2.2 &       25.7$\pm$ 31.1 &        83.9$\pm$ 82.8 &         3.6$\pm$  3.3&         0.9 &      5.9 &    ITA  \\   
	155&    138.2$\pm$ 27.0 &      362.1$\pm$ 79.7 &        7.5$\pm$  1.7 &       33.6$\pm$  9.9 &       210.7$\pm$120.9 &         1.5$\pm$  0.7&         1.2 &      8.3 &    ITA  \\   
	156&     43.0$\pm$  7.1 &      198.3$\pm$ 41.2 &       11.7$\pm$  2.5 &       34.8$\pm$  9.4 &       280.8$\pm$142.6 &         6.5$\pm$  2.8&         0.8 &      5.5 &    ITA  \\   
	157&     65.8$\pm$ 10.8 &      180.8$\pm$ 48.9 &        6.7$\pm$  1.9 &       64.0$\pm$ 21.3 &        84.7$\pm$ 38.6 &         1.3$\pm$  0.5&         1.0 &      6.6 &    ITA  \\   
	158&     64.7$\pm$ 22.8 &      340.5$\pm$121.2 &        6.1$\pm$  1.5 &       17.8$\pm$  8.7 &       132.4$\pm$ 70.9 &         2.0$\pm$  1.0&         0.8 &      5.5 &    ITA  \\   
	159&    148.3$\pm$ 11.9 &      518.9$\pm$ 52.3 &        9.4$\pm$  1.5 &       17.5$\pm$  2.3 &       478.5$\pm$168.8 &         3.2$\pm$  0.9&         0.9 &      5.9 &    ITA  \\   
	160&    267.6$\pm$ 29.1 &      319.0$\pm$ 56.4 &        5.5$\pm$  0.8 &       83.7$\pm$ 18.3 &        99.8$\pm$ 39.8 &         0.4$\pm$  0.1&         3.0 &     20.1 &    Arm  \\   
	161&     48.4$\pm$  7.4 &      109.2$\pm$ 26.2 &        8.1$\pm$  2.8 &      129.1$\pm$ 38.5 &        74.3$\pm$ 52.3 &         1.5$\pm$  0.9&         0.9 &      6.1 &    ITA  \\   
	162&     32.0$\pm$  9.7 &      255.6$\pm$ 98.4 &        5.2$\pm$  2.2 &       15.6$\pm$  7.8 &        72.5$\pm$ 49.8 &         2.3$\pm$  1.4&         0.9 &      6.2 &    ITA  \\   
	163&     30.9$\pm$ 12.0 &      293.7$\pm$123.1 &        2.6$\pm$  1.9 &       11.4$\pm$  6.5 &        20.4$\pm$ 35.5 &         0.7$\pm$  0.9&         1.0 &      7.1 &    ITA  \\   
	164&     64.1$\pm$ 27.7 &      618.0$\pm$122.1 &        6.4$\pm$  3.1 &        5.3$\pm$  2.2 &       267.1$\pm$302.6 &         4.2$\pm$  4.0&         0.8 &      5.4 &    ITA  \\   
	165&     73.0$\pm$ 30.5 &      321.7$\pm$ 95.5 &        2.9$\pm$  1.2 &       22.4$\pm$ 10.6 &        27.8$\pm$ 22.1 &         0.4$\pm$  0.3&         0.8 &      5.2 &    ITA  \\       
	\hline
	\caption{GMA Properties in M100. Column 10 lists the location of the GMAs. The GMAs are identified from the feathered (12m$+$ACA) spectral data cube using CPROPS.  Abbreviation of CNR and ITA in the column 10 stand for the circumnuclear ring and inter-arm regions, respectively.}
	\label{Tab_gmas}
\end{longtable*}


\end{document}